\documentclass[usenatbib]{mn2e}
\usepackage{epsfig}

\newcommand{\etal}{et~al.} 
\newcommand{\ionhy}{H{\sc ii} }
\newcommand{\UCHII}{UCH{\sc ii} }
\newcommand{\degrees}{$^\circ$}
\newcommand{\kms}{$\mbox{km~s}^{-1}$ }
\newcommand{\kmsns}{$\mbox{km~s}^{-1}$}

\newcommand{\vsfig}[2]           % Single FIGure (put one figures in the
{
  \begin{center}
    \begin{minipage}[t]{0.05\textwidth}
      {\footnotesize \raisebox{40mm}{(#2)}}
    \end{minipage}
    \begin{minipage}[t]{0.42\textwidth}
      \psfig{file=./#1.ps,height=0.95\textwidth,angle=0}
    \end{minipage}
    \hfill
  \end{center}
}

\newcommand{\specdfig}[2]        % Double FIGures (put two figures  
{
   \begin{center}
     \begin{minipage}[t]{0.45\textwidth}
         \psfig{file=eps/#1.eps,height=0.65\textwidth,width=1\textwidth,angle=0}
     \end{minipage}
     \hfill
     \begin{minipage}[t]{0.45\textwidth}
         \psfig{file=eps/#2.eps,height=0.65\textwidth,width=1\textwidth,angle=0}
     \end{minipage}
   \end{center}
}

\newcommand{\specsfig}[1]        % Single FIGure (put one figure of  
{
   \begin{center}
     \begin{minipage}[t]{0.45\textwidth}
         \psfig{file=eps/#1.eps,height=0.65\textwidth,width=1\textwidth,angle=0}
     \end{minipage}
   \end{center}
}

\newcommand{\boxfig}[1]        % Single FIGure (put one figure of  
{
   \begin{center}
     \begin{minipage}[t]{0.46\textwidth}
         \psfig{file=eps/#1.eps,height=0.45\textwidth,angle=0}
     \end{minipage}
   \end{center}
}

\newcommand{\twofig}[2]        % Double FIGures (put two figures  
{
   \begin{center}
     \begin{minipage}[t]{0.5\textwidth}
         \psfig{file=eps/#1.eps,height=0.95\textwidth}
     \end{minipage}
     \hfill
     \begin{minipage}[t]{0.5\textwidth}
         \psfig{file=eps/#2.eps,height=0.95\textwidth}
     \end{minipage}
   \end{center}
}

\begin{document}

\title[Quantifying high-mass star formation evolutionary schemes]{12.2-GHz methanol masers towards 1.2-mm dust clumps: Quantifying high-mass star formation evolutionary schemes}
\author[Breen \etal]{S.\ L. Breen,$^{1,2}$\thanks{Email: Shari.Breen@utas.edu.au} S.\ P. Ellingsen,$^1$ J.\ L. Caswell$^2$ \& B.\ E. Lewis$^1$\\
  \\
  $^1$ School of Mathematics and Physics, University of Tasmania, Private Bag 37, Hobart, Tasmania 7001, Australia\\
  $^2$ Australia Telescope National Facility, CSIRO, PO Box 76, Epping, NSW 1710, Australia}

 \maketitle
  
 \begin{abstract}
  
We report the results of a search for 12.2-GHz methanol maser emission, targeted towards 113 known 6.7-GHz methanol masers associated with 1.2-mm dust continuum emission. Observations were carried out with the Australia Telescope National Facility (ATNF) Parkes 64-m radio telescope in the period 2008 June 20 - 25. We detect 68 12.2-GHz methanol masers with flux densities in excess of our 5-$\sigma$ detection limit of 0.55~Jy, 30 of which are new discoveries. This equates to a detection rate of 60 per cent, similar to previous searches of comparable sensitivity. 

We have made a statistical investigation of the properties of the 1.2-mm dust clumps with and without associated 6.7-GHz methanol maser and find that 6.7-GHz methanol masers are associated with 1.2-mm dust clumps with high flux densities, masses and radii. 
We additionally find that 6.7-GHz methanol masers with higher peak luminosities are associated with less dense 1.2-mm dust clumps than those 6.7-GHz methanol masers with lower luminosities. We suggest that this indicates that more luminous 6.7-GHz methanol masers are generally associated with a later evolutionary phase of massive star formation than less luminous 6.7-GHz methanol maser sources. 

Analysis of the 6.7-GHz associated 1.2-mm dust clumps with and without associated 12.2-GHz methanol maser emission shows that clumps associated with both class~II methanol maser transitions are less dense than those with no associated 12.2-GHz methanol maser emission. Furthermore, 12.2-GHz methanol masers are preferentially detected towards 6.7-GHz methanol masers with associated OH masers, suggesting that 12.2-GHz methanol masers are associated with a later evolutionary phase of massive star formation. 

We have compared the colours of the GLIMPSE point sources associated with the maser sources in the following two subgroups: 6.7-GHz methanol masers with and without associated 12.2-GHz methanol masers; and 6.7-GHz methanol masers with high and and those with low peak luminosities. There is little difference in the nature of the associated GLIMPSE point sources in any of these subgroups, and we propose that the masers themselves are probably much more sensitive than mid-infrared data to evolutionary changes in the massive star formation regions that they are associated with.

We present an evolutionary sequence for masers in high-mass star formation regions, placing quantitative estimates on the relative lifetimes for the first time.

\end{abstract}


\section{Introduction}

Interstellar
masers are one of the most readily observed signposts of star
formation, particularly emission from the hydroxyl (OH), water (H$_2$O) and methanol (CH$_3$OH)
molecules as they are relatively common and intense. Because these masers arise at radio frequencies, they are not affected by the dense optical-obscuring gas and dust present at the early stages of massive star formation thereby allowing us to probe these stars at the earliest stages.  The masers are powerful probes of the kinematics of the star formation regions, but we are only recently beginning to be able
to use the masers to investigate aspects such as the physical conditions through comparison of observational data and maser pumping theories \citep{Cragg01,Sutton01}. The different maser species favour different physical conditions and are thought to trace different evolutionary phases of massive star formation \citep{Ellingsen07}.  Since many sources show emission from multiple maser transitions or species there must be significant overlap for the evolutionary phase traced by the most common types of masers. 

Interstellar masers have been detected towards more than a thousand star formation regions in our Galaxy from transitions of methanol, hydroxyl and water. Transitions of class II methanol masers have some advantages over water and OH masers as they exclusively trace sites of massive star formation \citep{Minier03,Xu08} while water and OH are known to also be associated with other astrophysical objects such as evolved stars and supernova remnants. Two of the strongest maser species found towards star formation regions are transitions of methanol at 6.7- and 12.2-GHz, both of which are known to trace an early evolutionary stage of massive star formation \citep{Ellingsen06}. Complementary observations of these two strong methanol maser transitions are especially useful in probing the physical conditions of the environments in which they arise, as they are known to typically be co-spatial \citep{Menten92,Norris93}.

To date there have been several searches for 12.2-GHz methanol maser emission towards 6.7-GHz methanol maser emission \citep*[e.g.][]{Caswell95b,Blas04}. \citet{Caswell95b} targeted their search for 12.2-GHz methanol masers towards 238 6.7-GHz methanol masers that had been detected
towards sites of OH masers and achieved a detection rate of 55 per cent. \citet{Blas04} directed their observations at 12.2-GHz towards 6.7-GHz methanol masers that were observed towards a
sample of {\em IRAS} (Infrared Astronomy Satellite) selected sources \citep{Szy00} as well as 6.7~GHz methanol masers that were detected in a blind search that had relatively poor positional accuracy and sensitivity \citep{Szy02}.

Searches such as those by \citet{Caswell95b} and \citet{Blas04} have shown
that 12.2-GHz methanol maser emission is only rarely brighter than the
associated 6.7-GHz methanol maser emission. This coupled with the fact
that there have been no serendipitous detections of 12.2-GHz masers
without a 6.7-GHz counterpart mean that it is unlikely that an
unbiased Galactic survey would uncover many, if any, more 12.2-GHz methanol masers than would be yielded from a search targeted towards 6.7-GHz methanol maser sources.

Recently, a small subset of the strong 6.7-GHz methanol masers from the sources observed by \citet{Hill05} at 1.2-mm, were searched for the presence of 12.2-GHz methanol masers with
the University of Tasmania 26~m Mount Pleasant radio telescope \citep{Lewis07}. \citet{Lewis07} detected 17 12.2-GHz methanol masers towards the 27 sites searched with a detection limit of $\sim$4~Jy. \citet{Hill05} observed 131 regions that were suspected of undergoing massive star formation, using the presence of previously identified 6.7-GHz methanol masers and/or an ultracompact \ionhy region to select these regions. A total of 404 1.2-mm dust clumps were identified, with 113 of these associated with 6.7-GHz methanol masers, 35 of which are associated with radio continuum emission \citep{Walsh98}.

Statistical analysis by \citet{Lewis07} of the properties of 1.2-mm dust clumps associated with 6.7-GHz methanol maser emission with and without 12.2-GHz methanol masers showed that those 1.2-mm dust clumps with both 6.7-GHz and 12.2-GHz were less massive and less dense than the 1.2-mm clumps that were associated with only 6.7-GHz methanol maser emission.  Comparison between the 6.7-GHz methanol masers with detected 12.2-GHz with those with no detectable 12.2-GHz maser emission, showed that 12.2-GHz methanol masers were preferentially found at those 6.7-GHz methanol maser sites with associated OH emission (to a better than 95 per cent confidence). As OH masers are expected to trace a generally later stage of star formation \citep{FC89,Cas97} than 6.7-GHz methanol masers. As the sample size of the \citet{Lewis07} observations was small, was subject to a 6.7-GHz flux density bias and had relatively poor sensitivity, the aim of the current work was to confirm these results by testing a large sample with a much lower 12.2-GHz detection limit.

Here we present a targeted search for 12.2-GHz methanol masers, with a greatly increased sample size and sensitivity (c.f. \citet{Lewis07}), towards 113 known 6.7-GHz methanol masers which have been previously targeted for 1.2-mm dust continuum emission \citep{Hill05}. Statistical analysis of the 1.2-mm dust clumps devoid of either methanol maser transition and those associated with 12.2-GHz and/or 6.7-GHz methanol maser sources has been carried out, with a particular emphasis on the investigation of the relative evolutionary stage each class of source is tracing.

\section{Observations \& Data Reduction}\label{section:obs}

We have searched 113 6.7-GHz methanol masers, towards which \citet{Hill05} observed 1.2-mm dust clump continuum emission, for the presence of associated 12.2-GHz methanol maser emission. Observations were carried out with the Australia Telescope National Facility (ATNF) Parkes 64-m radio telescope during 2008 June 20 - 25. Follow-up observations of two sources were made on 2008 December 7 using an identical set-up to the observations carried out in 2008 June. The observations were made with the Ku-band receiver which detected two orthogonal linear polarizations and had typical system temperatures of 205 and 225 Jy for the respective polarizations throughout the observations. The Parkes multibeam correlator was configured to record 8192 channels over 16-MHz for each of the recorded linearly polarized signals. This configuration yielded a useable velocity coverage of $\sim$290 \kms and a spectral resolution of 0.08 \kmsns, after Hanning smoothing. The Parkes radio telescope has rms pointing errors of $\sim$10 arcsec and at 12.2-GHz the telescope has a half power beam width of 1.9 arcmin.

All sources were observed at a fixed frequency of 12178-MHz
(i.e. with no Doppler tracking), which alleviated the requirement for
a unique reference observation to be made for each of the source
positions. Instead, all the spectra collected each day were combined
to form a sensitive reference bandpass by taking the median value of
each channel.  The noise contribution to the
quotient spectrum from such a reference is negligible and means that
the theoretical radiometer noise is achieved for each observation. 

As the bandpass is not constant throughout an observing session, observing in this mode introduced ripples into the quotient spectra. This was overcome by subtracting a running median over 100 channels from the baseline. As the maser lines are much narrower than 100 channels this has little effect on emission spectra but may have an adverse effect on absorption features which are of a comparable width. Our ability to detect absorption is therefore reduced and in most cases where we do detect absorption it appears narrower and weaker than it would have been if we had employed a more traditional, but less efficient, observing method. In order to determine the absorption characteristics of the sources, spectra where absorption features were detected were boxcar smoothed over 20 channels, decreasing our spectra resolution to 0.26 \kms and resulted in a typical rms noise of $\sim$0.04~Jy. 

Data were reduced using ASAP (ATNF Spectral Analysis Package). Alignment of velocity channels was carried out during processing. Absolute flux density calibration was achieved by observing PKS B1934--638 each day which has an assumed flux density of 1.825 Jy at 12178-MHz \citep{Sault03}. Each target was observed for 10 minutes, and once the polarizations were averaged yielded a typical rms noise of 0.11~Jy corresponding to a 5-$\sigma$ detection limit of 0.55~Jy. For weak sources (of comparable strength to our detection limit), individual polarizations were inspected as a further check on the reliability of the detection. The adopted rest frequency was 12.178597 GHz \citep{Muller04}.

\section{Results}
Our search, for 12.2-GHz methanol masers, targeted 113 6.7-GHz methanol maser sites and resulted in the detection of 68 12.2-GHz methanol masers, equating to a detection rate of 60 per cent. A search of the literature \citep*[e.g.][]{Pestalozzi05,Blas04,Caswell95b} reveals that 30 of these are new detections. A number of the 12.2-GHz methanol masers that have been previously observed, particularly the stronger ones, were discovered prior to \citet{Blas04,Caswell95b} by others \citep[e.g.][]{Gay,Cas93}. However, direct comparison of the majority of these sources are complicated by large positional uncertainties, pointing offsets and the limited velocity ranges covered in the observations.

Table~\ref{tab:sources} summarizes information on the targeted 6.7-GHz methanol masers  \citep{Caswell09,Minier03,Walsh98,Walsh97,Caswell95a,Szy02}, associated 12.2-GHz methanol masers that we detect, along with the 1.2-mm dust clump \citep{Hill05} that each methanol maser site is associated with. For the majority of sources, precise positions for the 6.7-GHz methanol maser sources have been determined with the Australia Telescope Compact Array to an accuracy better than 1 arcsec. There are five instances where two methanol masers are associated with the same dust clump, therefore our observations targeted a total of 106 1.2-mm dust clumps. Spectra of the detected 12.2-GHz methanol masers are shown in Fig.~\ref{fig:spectra} and are presented in order of Galactic longitude except for where vertical alignment of nearby sources was necessary to highlight where features of a source were also detected at nearby positions. Comments on some individual sources can be found in Section~\ref{Section:sources}. The absorption features that we detect are presented in Table~\ref{tab:abs}.

\begin{table*}
  \caption{Targeted 6.7-GHz methanol masers, detected 12.2-GHz methanol masers and associated 1.2-mm dust clump that they are associated with. Column 1 gives the Galactic coordinates for each methanol maser source, derived from the position of the 6.7-GHz methanol maser, columns 2 and 3 give the 6.7-GHz methanol maser right ascension and declination. References for each 6.7-GHz methanol maser position are represented by the superscript number at the end of the source name where 1: \citet{Caswell09}, 2: \citet{Minier03}, 3: \citet{Walsh98}, 4: \citet{Walsh97}, 5: \citet{Caswell95a}, 6: \citet{Szy02}, $\alpha$: James Caswell, private communication; see also note in comments on sources for G\,034.246+0.134 and G\,035.025+0.350.  References for previously detected 12.2-GHz methanol masers are listed after the 6.7-GHz methanol maser references and are separated by a comma and contained within brackets to avoid confusion, here 7: \citet{Caswell95b}, 8: \citet{Blas04}, $^{*}$: new detection. Columns 4 - 10 list the 6.7-GHz peak flux density (Jy), peak velocity (\kmsns), velocity range (\kmsns) (nb. where the source reference is \citet{Walsh98} the velocity range has been taken from \citet{Walsh97} where available and \citet{Pestalozzi05} if not), 12.2-GHz peak flux density (Jy), peak velocity (\kmsns), velocity range (\kmsns) and integrated flux density (Jy \kmsns). The 5-$\sigma$ detection limit and velocity range covered at 12.2-GHz is presented for sources where we detect no 12.2-GHz emission, in these cases the 5-$\sigma$ detection limit is presented in column 7 and is preceded by a $<$, and the velocity ranges observed are listed in column 9. $^{\beta}$ indicates a source that was previously observed by \citet{Caswell95b} but we do not detect. Column 11 indicates if the methanol maser sites have been searched for the presence of an associated OH maser by \citet{Caswell98}, -- indicates that no suitable data were found, n indicates that the site has been searched but no source was detected and y indicates that there is an OH maser within 2 arcsec of the 6.7-GHz methanol maser. The dust clump \citep{Hill05} that the methanol masers are coincident with are listed in column 12, a $^{\#}$ after the source name indicates sources that were not included in the statistical analysis, for details see Section~\ref{Section:stats}.}
  \begin{tabular}{lllrrcrrcllll} \hline
  \multicolumn{1}{c}{\bf Methanol maser} & {\bf RA}  & {\bf Dec} & {\bf S$_{6.7}$} & {\bf Vp$_{6.7}$} & {\bf Vr$_{6.7}$ } & {\bf S$_{12.2}$} & {\bf Vp$_{12.2}$} & {\bf Vr$_{12.2}$} & {\bf I$_{12.2}$} & {\bf OH} &  {\bf Dust clump}\\
    \multicolumn{1}{c}{\bf ($l,b$)}&  {\bf (J2000)} & {\bf (J2000)}  &{\bf (Jy)} & & &{\bf (Jy)}&& & & &{\bf ($l,b$)}  \\	
      \multicolumn{1}{c}{\bf (degrees)}  & {\bf (h m s)}&{\bf ($^{o}$ $'$ $``$)}&  &&&&&&&&{\bf (degrees)}\\  \hline \hline

G\,189.778+0.345$^{1}$		& 06 08 35.28		& +20 39 06.7		&	15	&	6	&	2,6	&$<$0.75	&		& --163,125		&		&	--	&	G\,189.78+0.34		\\
G\,192.600--0.048$^{1,(*)}$		& 06 12 53.99		& +17 59 23.7		& 	72	&	5.0	&	2,6	&	0.47	&	3.6	&	3.4,3.7	&	0.1	&	--	&	G\,192.60--0.05	\\
G\,206.543--16.355$^{2}$	& 05 41 44.15		& --01 54 44.9		&	1.5	&	--1	&		&$<$0.6	&		&--165,123	&		&	--	&	G\,206.54--16.35	\\
G\,213.705--12.597$^{1,(7)}$	& 06 07 47.87		& --06 22 57.0		& 	337	&	12	& 	8,13	&	14	&	12.7	&	12.2,13.3	&	6.0	&	y	&	G\,213.61--12.6	\\
G\,259.939--0.041$^{3}$		& 08 35 31.086		& --40 38 23.96		& 	2.5	&	--1.3	&		&$<$0.55	&		& --180,108	&		&	n	&	G\,259.94--0.04$^{\#}$	\\
G\,269.153--1.128$^{1}$		& 09 03 33.46		& --48 28 02.6		&	1.6	& 	16	&	7,16	&$<$0.75	&		&	--179,109	&		&	n	&	G\,296.15--1.13	\\
G\,269.456--1.467$^{3,(*)}$		& 09 03 14.848		& --48 55 11.25		& 	4.5	&	56	&		&	1.4	&	56.0	& 55.9,56.3	&	0.4	&	n	&	G\,296.45--1.47	\\
G\,270.255+0.835$^{1}$		& 09 16 41.51		& --47 56 12.1		& 	0.6	&	3.9	&	3,5	&$<$0.75	&		&	--179,109	&		&	n	&	G\,270.25+0.84		 \\
G\,284.352--0.419$^{1,(7)}$	& 10 24 10.89		& --57 52 38.8		&	1.7	&3.3		&	3,11	&$<$0.80$^{\beta}$	&		& --176,112		&		&	n	&	G\,284.35--0.42$^{\#}$	\\
G\,287.371+0.644$^{1,(*)}$		& 10 48 04.40		& --58 27 01.7 	&	80	&	--1.8	&	--3,0	&	49	& --2.0	&--2.5,--1.0	&	20.2	&	y	&	 G\,287.37+0.65		\\
G\,290.374+1.661$^{1}$		& 11 12 18.10		& --58 46 21.5		&	0.6	& --24.2	&--28,--22	&$<$0.6 	&		&	--174,115	&		&	y	&	G\,290.37+1.66$^{\#}$		    \\
G\,290.411--2.914$^{3,(*)}$		& 10 57 33.987		& --62 59 03.13		& 	2.4	&--16.2	&--17,--15	&	1.5	&--16.1	& --16.2,--15.8	& 0.4		&	n	&	G\,290.40--2.91	\\
G\,291.274--0.709$^{1,(7)}$	& 11 11 53.35		& --61 18 23.7		& 100	&--29.6	&--31,--28	&	3.1	& --29.1	& --30.1,--28.7	&	2.4	&	y	&	G\,291.27--0.70	\\
G\,291.579--0.431$^{1}$		& 11 15 05.76		& --61 09 40.8		&	0.7	&14.5	&11,16	&$<$0.6  	&		& --173,115	&		&	y	&	G\,291.58--0.53$^{\#}$	\\
G\,291.582--0.435$^{1}$		& 11 15 06.61		& --61 09 58.3		&	1.7	&10.5	&	8,11	&$<$0.6	&		&--173,115	&		&	n	&	G\,291.58--0.53$^{\#}$	\\
G\,293.827--0.746$^{3}$		& 11 32 05.606		& --62 12 25.43		& 3.6		& 36.7	&35,39	&$<$0.56	&		&--173,115			&		&	n	&	G\,293.82--0.74	\\
G\,293.942--0.874$^{4,(*)}$		& 11 32 42.091		& --62 21 47.51		& 3.6		&	41.0	&		&	2.1	&	41.1	& 40.9,41.3	&	0.6	&	n	&	G\,293.942--0.876		\\ 
G\,294.511--1.621$^{1}$		& 11 35 32.25		& --63 14 43.2		&	12	&--10.2	&--14,--9	&$<$0.55	&		& --172,116	&		&	y	&	G\,294.52--1.60  	\\
G\,294.990--1.719$^{1}$		& 11 39 22.88		& --63 28 26.4		&	18	&--12.3	&--13,--11	&$<$0.55	&		& --172,116	&		&	n	&	G\,294.989--1.720	\\
G\,298.262+0.739$^{3}$		& 12 11 47.655		& --61 46 21.14		& 11		&--30.2	&--31,--29	&$<$0.55	&		& --172,116	&		&	y	&	G\,298.26+0.7		 \\
G\,299.013+0.128$^{1}$		& 12 17 24.60		& --62 29 03.7		&	7	&18.4	&18,20	&$<$0.56	&		& --171,117	&		&	y	&	G\,299.02+0.1		 \\
G\,300.504--0.176$^{1}$		& 12 30 03.58		& --62 56 48.7		&4.7		& 7.5		&	4,11	&$<$0.55	&		&--170,118	&		&	y	&	G\,300.51--0.1		\\
G\,301.136--0.226$^{1}$		& 12 35 35.14		& --63 02 32.6		& 1.2		&--39.8	&--41,--37	&$<$0.52	&		& --146,142	&		&	y	&	G\,301.14--0.2		\\
G\,302.032--0.061$^{1}$		& 12 43 31.92		& --62 55 06.7		& 11		&--35.3	&--43,--33	&$<$0.6	&		& --170,118			&		&	n	&	G\,302.03--0.06	\\
G\,305.199+0.005$^{1,(*)}$		& 13 11 17.20		& --62 46 46.0		&	2.3	& --42.8	&--45,--38	&	0.7	& --42.6	& --42.7,--41.4	&	0.1	&	n	&	G\,305.192--0.006	\\
G\,305.200+0.019$^{1,(*)}$   	& 13 11 16.93		& --62 45 55.1		& 44		& --33.1	&--38,--29	&	10.6	& --31.8	& --33.2,--31.6	&	8.8	&	y	&	G\,305.200+0.02	\\
G\,305.202+0.208$^{1,(7)}$	& 13 11 10.49		& --62 34 38.8  & 20 & --43.9 & --47,--43 & 3.5 & --43.9	& --45.0,--43.6	&	1.8	&	y	&	G\,305.21+0.21		 \\ 	
G\,305.208+0.206$^{1,(7)}$		& 13 11 13.71		& --62 34 41.4		& 320	& --38.3	&--42,--34	&	1.7	& --36.5	&--40.5, --35.7	&	0.9	&	y	&	G\,305.21+0.21		 \\ 
G\,305.248+0.245$^{1}$		& 13 11 32.47		& --62 32 09.1		& 4		& --32.0	&--36,--28	&$<$0.60&		&--169,119	&		&	n	&	G\,305.248+0.245	 \\
G\,305.362+0.150$^{1}$		& 13 12 35.86		& --62 37 17.9		& 3		& --36.5	&--38,--35	&$<$0.55	&		&--169,119	&		&	y	&	G\,305.361+0.151	\\
G\,305.366+0.184$^{1,(7)}$		& 13 12 36.74		& --62 35 14.7		& 2.5		& --33.8	&--35,--33	&	3.3	& --34.1	& --34.4,--33.6	& 	1.6	&	n	&	G\,305.362+0.185	 \\
G\,305.563+0.013$^{\alpha,(*)}$		& 13 14 26.90    	& --62 44 29.4		&  4.5		& --37.3	&--42,--32	&	1.6	& --37.1	& --41.6,--35.7	&	1.4	&	n	&	G\,305.552+0.013	 \\ 
G\,305.799--0.245$^{1}$		& 13 16 43.23		& --62 58 32.9		&	0.7	& --39.5	&--40,--36	&$<$0.55		&		& --168,120		&		&	y	&	G\,305.81--0.25	\\
G\,306.322--0.334$^{1}$		& 13 21 23.01		& --63 00 29.5		& 	1.2	& --24.4	&--25,--22	& $<$0.55	&		&--168,120	&		&	y	&	G\,306.33--0.3		\\
G\,309.921+0.479$^{1,(7)}$		& 13 50 41.78		& --61 35 10.2	&	635	& --59.8	&--65,--54	& 	179	& --59.7	& --61.3,--57.7	& 108.5	&	y	&	G\,309.92+0.4		\\
G\,318.948--0.196$^{1,(7)}$		& 15 00 55.39		& --58 58 52.8		&	690	& --34.7	&--39,--31	&	140	& --34.4	& --35.5,--32.0	& 91.0	&	y	&	G\,318.92--0.68	 \\
G\,323.740--0.263$^{1,(7)}$		& 15 31 45.45   	& --56 30 50.1		& 	3000	&--51.1	&--58,--45	&	420	& --48.7	& --52.6,--47.8	& 785.1	&	y	&	G\,323.74--0.3		\\
G\,330.953--0.182$^{1}$		& 16 09 52.37		& --51 54 57.6		& 	7	& --87.6	&--90,--87	& $<$0.55	&		& --156,132			&		&	n	&	G\,330.95--0.18	\\
G\,331.278--0.188$^{1,(7)}$		& 16 11 26.59		& --51 41 56.7		& 	190	&--78.2	&--87,--77	&	76	&  --78.7	& --83.1,--77.7	& 58.8	&	y	&	G\,331.28--0.19$^{\#}$	\\
G\,332.653--0.621$^{1}$		& 16 19 43.51		& --51 03 39.9		& 7.1		& --50.6	&--52,--49	& $<$0.54	&		& --130,158	&		&	n	&	G\,332.648--0.606		\\
G\,332.726--0.621$^{1,(*)}$		& 16 20 03.00		& --51 00 32.5		& 2.7		&--49.5	&--57,--44	&	0.5	& --47.3	& --47.4,--47.2	& 0.1		&	y	&	G\,332.73--0.62		\\
G\,000.212--0.001$^{1,(*)}$		& 17 46 07.63		& --28 45 20.9		& 3.5		& 49.2	&41,50	&	0.8	& 49.5	& 49.1,49.7	& 0.5		&	n	&	G\,0.21--0.00			\\
G\,000.315--0.201$^{1,(*)}$		& 17 47 09.13		& --28 46 15.7		& 41.2	& 18		&14,27	&	2.4	& 18.3	& 17.8,18.5	& 1.0		&	n	&	G\,0.32--0.20		\\
G\,000.316--0.201$^{1}$		& 17 47 09.33		& --28 46 16.0		& 1.3		& 21		&20,22	&	$<$0.55	& 	& --120,168	&		&	n	&	G\,0.32--0.20		 \\

 \hline
\end{tabular}\label{tab:sources}
\end{table*}
\clearpage 

\begin{table*}\addtocounter{table}{-1}
  \caption{-- {\emph {continued}}}
  \begin{tabular}{lllrrcrrcllll} \hline
  \multicolumn{1}{c}{\bf Methanol maser} & {\bf RA}  & {\bf Dec} & {\bf S$_{6.7}$} & {\bf Vp$_{6.7}$} & {\bf Vr$_{6.7}$ } & {\bf S$_{12.2}$} & {\bf Vp$_{12.2}$} & {\bf Vr$_{12.2}$} & {\bf I$_{12.2}$} & {\bf OH} &  {\bf Dust clump}\\
    \multicolumn{1}{c}{\bf ($l,b$)}&  {\bf (J2000)} & {\bf (J2000)}  &{\bf (Jy)} & & &{\bf (Jy)}&& & & &{\bf ($l,b$)}  \\	
      \multicolumn{1}{c}{\bf (degrees)}  & {\bf (h m s)}&{\bf ($^{o}$ $'$ $``$)}&  &&&&&&&&{\bf (degrees)}\\  \hline \hline

G\,000.496+0.188$^{1,(*)}$		& 17 46 03.96		& --28 24 52.8		& 10		&0.8		& --12,2	&	18	& 1.1		& --9.5,1.5		& 15.6	&	y	&	G\,0.49+0.19		\\ 
G\,000.546--0.852$^{1,(7)}$		& 17 50 14.35		& --28 54 31.1		& 50		&	13.8	&8,20	&	3.9	& 18.5	& 16.1,19		& 2.1		&	y	&	G\,0.55--0.85			\\
G\,000.836+0.184$^{1}$		& 17 46 52.86		& --28 07 34.8		& 8.1		&	3.5	&	2,4	& $<$0.52	&		&--115,173	&		&	n	&	G\,0.83+0.18			\\
G\,001.147--0.124$^{3}$		& 17 48 48.467		& --28 01 11.81		& 4		&--20.8	& --21,--15&$<$0.51	&		& --115,173	&		&	n	&	G\,1.14--0.12\\
G\,002.536+0.198$^{1,(*)}$		& 17 50 46.47		& --26 39 45.3		& 40		&	3.2	&	2,20	&	3.7	& 7.0		& 3.4,7.2		& 2.6		&	n	&	G\,2.54+0.20		 \\
G\,005.900--0.430$^{1}$		& 18 00 40.86		& --24 04 20.8		& 5.3		&	10	&4,11	& $<$0.55	&		& --113,175			&		&	n	&	G\,5.90--0.42			\\
G\,006.610--0.082$^{1,(*)}$		& 18 00 54.03		& --23 17 03.1		& 10.8	&	0.7	& 0,1		&	 10.7	& 0.9		& 0.5,1.4		&	4.7	&	n	&	G\,6.60--0.08$^{\#}$		 \\
G\,008.139+0.226$^{1}$		& 18 03 00.75		& --21 48 09.9		&	3.5	& 20		&19,21	& $<$0.55&		& --112,176			&		&	n	&	G\,8.13+0.22		  \\
G\,008.669--0.356$^{1,(*)}$		& 18 06 18.99		& --21 37 32.2		& 8.5		&	39.3	&39,40	&	0.3	& 39.5	& 39.4,39.6	&	0.05	&	y	&	G\,8.68--0.36		\\
G\,008.683--0.368$^{1,(7)}$		& 18 06 23.49		& --21 37 10.2		&	70	&	42.9	&40,46	&	8.4	&44.2	& 41.0,45.5	&	11.6	&	y	&	G\,8.686--0.366	 	\\
G\,009.619+0.193$^{1,(*)}$		& 18 06 14.92		& --20 31 44.3		& 	72	&	5.5	&	5,7	&	0.4	& 5.8		& 5.7,5.9		& 	0.1	&	y	&	G\,9.62+0.19			\\ 
G\,009.621+0.196$^{1,(7)}$		& 18 06 14.67		& --20 31 32.4	& 5000	&	1.3	&	--4,9	&	401	& 1.4		& --3.0,5.8		&  202.8	&	y	&	G\,9.62+0.19			\\
G\,009.986--0.028$^{1,(*)}$		& 18 07 50.12		& --20 18 56.5		&	28	&	47.1	&40,52	&	1.0	& 47.1	& 44.3,47.3	& 0.5		&	n	&	G\,9.99--0.03		\\
G\,010.287--0.125$^{1}$		& 18 08 49.36		& --20 05 59.0		& 	27	&	5.0	&	4,6	&$<$0.52	&		&  --111,178			&		&	n	&	G\,10.284--0.126	 	\\
G\,010.299--0.146$^{1}$		& 18 08 55.54		& --20 05 57.5		& 	3.5	&	20	&19,21	&$<$0.54	&		& --111,178			&		&	n	&	G\,10.29--0.14			\\
G\,010.323--0.160$^{1,(*)}$	& 18 09 01.46	& --20 05 07.8		& 	126	&	10	&4,14	&	2.3	& 10.3	& 5.1,12.6		& 1.8		&	n	&	G\,10.32--0.15		 \\
G\,010.444--0.018$^{1,(7)}$	& 18 08 44.88	& --19 54 38.3		& 14.8	&	73.2	& 68,79	&	6.5	& 72.0	& 68.0,73.7	& 10.1	&	y	&	G\,10.44--0.01			\\
G\,010.473+0.027$^{1,(7)}$	& 18 08 38.20	& --19 51 50.1		& 120	&	75.0	&58,77	&	12.4	& 75.1	& 73.7,76.7	& 9.1		&	y	&	G\,10.47+0.02		\\
G\,010.480+0.033$^{1}$	& 18 08 37.88	& --19 51 16.0	&	9.7	&	65.0	&	58,66	& $<$0.55	&		&	--111,178		&		&	y	&	G\,10.47+0.02		\\
G\,010.627--0.384$^{1,(7)}$	& 18 10 29.22	& --19 55 41.1	& 	3.1	&	4.6	& --6,7		&	1.4	&	4.6	&	4.0,6.6	&	0.8	&	n	&	G\,10.62--0.38			\\
G\,010.629--0.333$^{1}$	& 18 10 17.98	& --19 54 04.8	& 	4.2	&	--7.5	& 	--13,1	& $<$0.54	&		&--111,178	&		&	n	&	G\,10.62--0.33			\\
G\,011.497--1.485$^{1,(*)}$	& 18 16 22.13	& --19 41 27.1	& 	167	&	6.7	&	4,17		&	29	&	9.0	&	6.0,17.1	&	15.8	&	n	&	G\,11.49--1.48			\\
G\,011.904--0.141$^{1,(7)}$	& 18 12 11.44	& --18 41 28.6	& 	56	&	42.8	&	40,45	&	17	&	43.0	&	42.1,44.0	&	8.2	&	y	&	G\,11.903--0.140	 \\
G\,011.936--0.150$^{1}$	& 18 12 17.29	& --18 40 02.6	&	1.9	&	48.5	&	47,50	&$<$0.76	&		&--111,178		&		&	n	&	G\,11.93--0.14		\\
G\,011.936--0.616$^{1,(7)}$	& 18 14 00.89	& --18 53 26.6	&	41	&	32.2	&	30,44	&	4.3	&	39.8	&	31.5,42.4	&	4.4	&	n	&	G\,11.93--0.61		\\
G\,011.992--0.272$^{3}$	& 18 12 51.197&  --18 40 39.7&	0.6	&	59.5	&	59,61	&$<$0.75	&		& --110,179			&		&	n	&	G\,11.99--0.27		\\
G\,012.025--0.025$^{1,(7)}$	& 18 12 01.86	& --18 31 55.7	& 	85	& 107.7	& 105,112		&	14	&   108.6	&107.8,109.4	&	8.8	&	n	&	G\,12.02--0.03		\\
G\,012.202--0.120$^{1}$	& 18 12 42.93	&  --18 25 11.8& 	1.7	&	26.4	&	26,27	&$<$0.75	&		& --111,178		&		&	n	&	G\,12.18--0.12		\\
G\,012.209--0.102$^{1}$	& 18 12 39.92	& --18 24 17.9& 	9.2	&	19.8	&	16,22	&$<$0.75	&		&--111,178		&		&	n	&	G\,12.20--0.09		\\
G\,012.681--0.182$^{1,(7)}$	& 18 13 54.75	& --18 01 46.6	&	450	&	57.6	&	50,61	& 	8	&	57.1	& 55.6,60.3	&	7.6	&	y	&	G\,12.68--0.18		 \\
G\,012.889+0.489$^{1,(7)}$	& 18 11 51.39	& --17 31 29.6	& 	27	&	39.3	&	28,42	&	19	&	39.4	& 38.9,40.2	&	8.7	&	y	&	G\,12.88+0.48		 	\\
G\,012.909--0.260$^{1,(7)}$	& 18 14 39.53	& --17 52 00.0	& 	300	&	39.9	&	35,47	&	16	&	39.4	& 38.5,40.6	&	14.0	&	y	&	G\,12.90--0.26		\\
G\,014.60+0.01$^{3,(*)}$     	& 18 17 01.139& --16 14 38.88	&	1.2	&	23.2	&	22,28	&	0.6	&	24.7	&	24.6,24.7	&	0.1	&	n	&	G\,14.60+0.01		 \\
G\,015.034--0.677$^{1,(7)}$	& 18 20 24.78	& --16 11 34.6	& 	28	&	21.2	&	20,24	&	21	&	23.4	&	21.2,23.7	&	12.1	&	y	&	G\,15.03--0.67		 \\
G\,016.58--0.05$^{3,(7)}$      	& 18 21 09.132	& --14 31 48.69	& 	22	&	59.3	&	52,69	&	2.2	&	65.9	&	59.2,68.7	&	2.2	&	y	&	G\,16.58--0.05\\
G\,016.86--2.15$^{3}$      	& 18 29 24.411	& --15 16 04.12	& 	24	&	14.8	&	14,19	&$<$0.53	&		& --108,180		&		&	y	&	G\,16.86--2.15		 	\\ 
G\,019.609--0.234$^{\alpha}$	& 18 27 37.99	& --11 56 37.6	&	0.4	&	40.1	&	36,43	&	$<$0.53&		& --107,181		&		&	y	&	G\,19.607--0.234		\\
G\,019.612--0.134$^{3,(7)}$	& 18 27 16.524	& --11 53 38.49	&	18	&	56.2	&	50,61	& 	1.3	&	50.5	&	50.2,55.1	&	0.8	&	--	&	G\,19.61--0.1		 \\
G\,021.880+0.014$^{3}$	& 18 31 01.749	& --09 49 01.13	& 	5.5	&	20.6	&	17,22	&$<$0.52	&		&--106,182	&		&	--	&	G\,21.87+0.01		\\
G\,022.356+0.066$^{3,(*)}$	& 18 31 44.128	& --09 22 12.40	&	10.0	&	80.0	&	 77,88	&	1.0	&	81.4	&	81.4,81.5	&	0.1	&	--	&	G\,22.35+0.06		\\
G\,023.257--0.241$^{3,(*)}$	& 18 34 31.259	& --08 42 46.84	&	6.0	&	64.0	&	63,66	& 	0.9	&	64.9	&	63.8,65.9	&	0.7	&	--	&	G\,23.25--0.24		\\
G\,023.437--0.184$^{1,(7)}$	& 18 34 39.25 & --08 31 38.5&	45	&103.0	&	101,108	& 	3.3	& 104.0	& 103.6,105.3	&	1.4	&	y	&	G\,23.43--0.18		 \\
G\,023.440--0.182$^{1,(7)}$	& 18 34 39.18	& --08 31 25.3	& 	25	&	96.6	&	94,100	&	9	&	97.5	& 96.1,107.3	&	16.0	&	--	&	G\,23.43--0.18		\\
G\,024.790+0.083$^{3}$	& 18 36 12.571	& --07 12 11.31	& 	63	&113.3	& 	106,115	&$<$0.52	&		&--104,184	&		&	y	&	G\,24.78+0.08		 \\
G\,024.850+0.087$^{3,(*)}$	& 18 36 18.398	& --07 08 52.09	& 	24	& 109.7	&	109,115	&	1.8	&	110.1& 107.7,110.7	&	1.2	&	--	&	G\,24.84+0.08		 \\
G\,025.650+1.050$^{3}$	& 18 34 20.913	& --05 59 40.44	& 	90	&41.7	&	38,43	&$<$0.51	&		& --104,184			&		&	--	&	G\,25.65+1.04		 \\
G\,025.710+0.044$^{3,(8)}$	& 18 38 03.148	& --06 24 14.80	& 	355	&	95.5	&	89,101	&	144	&	95.6	& 89.3,100.4	& 107.7	&	--	&	G\,25.70+0.04		\\
G\,025.826--0.178$^{3,(*)}$	& 18 39 03.631	& --06 24 09.64	& 	36	&	91.5	&	89,94	&	25	&	90.9	& 90.0,92.1	&	14.7	&	--	&	G\,25.83--0.18		 \\
G\,028.146--0.005$^{1,(7)}$	&	18 42 42.59	&	 --04 15 36.5	& 	61	&	 101.2	&	 99,105	& 20	& 101.3	& 99.9,104.1	&	11.9	&	--	&	G\,28.14--0.00	 \\ 
G\,028.201--0.049$^{1}$	& 18 42 58.08	& --04 13 56.2	& 	3.5	&	98.9	&	94,100	&$<$0.52	& 		& --103,185			&		&	y	&	G\,28.20--0.04		 \\ 
G\,028.282--0.359$^{3,(*)}$	& 18 44 13.270	& --04 18 02.99&	1.1	&	40.4	&			&	0.7	&	41.0	&	40.8,41.0	&	0.1	&	--	&	G\,28.28--0.35		 \\
G\,028.305--0.387$^{3,(*)}$	& 18 44 21.988	& --04 17 38.42	&	66	&	81.0	& 	79,93	&	8	&	81.9	&	80.8,83.0	&	7.9	&	--	&	G\,28.31--0.38		\\
G\,029.865--0.043$^{3,(7)}$	& 18 45 59.588	& --02 44 57.60	&	12	& 101.2	&	95,105	& 	21	&	100.4& 99.9,102.1	& 10.1	&	--	&	G\,29.86--0.04		 \\
G\,029.956--0.016$^{3,(7)}$	& 18 46 03.763	& --02 39 19.69	& 	158	&	95.7	&	95,102	&	51	&	96.6	& 	91.2,98.5	&	45.0	&	--	&	G\,29.96--0.02B	\\
G\,029.979--0.047$^{3,(7)}$	& 18 46 12.974	& --02 38 58.05	& 	14	&	98.2	&	97,105	&	6.6	&	98.3	&	96.5,103.9& 	4.9	&	--	&	G\,29.978--0.050	\\
G\,030.591--0.042$^{3}$	& 18 47 18.886	& --02 06 07.00	&	0.6	&	43.2	&	36,49	&$<$0.55	&		& --102,186		&		&	--	&	G\,30.59--0.04		 \\
G\,030.704--0.068$^{3}$	& 18 47 36.801	& --02 00 48.99	& 	66	&	88.0	&	85,90	&$<$0.55	&		&--102,186			&		&	--	&	G\,30.705--0.065	 \\
G\,030.760--0.052$^{3,(7)}$	& 18 47 39.729	& --01 57 21.92 &	75	&	91.8	&	88,95 	&	1.7	&	91.3	&	90.7,91.8	&	0.9	&	--	&	G\,30.769--0.048	 \\
G\,030.818--0.057$^{3,(7)}$	& 18 47 46.984	& --01 54 19.61	&	10	&  101.2	&	91,110	&	0.5	& 108.1	&108.0,108.1	&	0.1	&	--	&	 G\,30.81--0.05		 \\
G\,030.898+0.162$^{3,(*)}$	& 18 47 09.133	& --01 44 10.29	&	40	& 101.8	&		 	&	14	& 101.7	& 99.3,102.1	&	13.5	&	--	&	G\,30.89+0.16		 \\
\hline
\end{tabular}
\end{table*}

\begin{table*}\addtocounter{table}{-1}
  \caption{-- {\emph {continued}}}
  \begin{tabular}{lllrrcrrcllll} \hline
  \multicolumn{1}{c}{\bf Methanol maser} & {\bf RA}  & {\bf Dec} & {\bf S$_{6.7}$} & {\bf Vp$_{6.7}$} & {\bf Vr$_{6.7}$ } & {\bf S$_{12.2}$} & {\bf Vp$_{12.2}$} & {\bf Vr$_{12.2}$} & {\bf I$_{12.2}$} & {\bf OH} &  {\bf Dust clump}\\
    \multicolumn{1}{c}{\bf ($l,b$)}&  {\bf (J2000)} & {\bf (J2000)}  &{\bf (Jy)} & & &{\bf (Jy)}&& & & &{\bf ($l,b$)}  \\	
      \multicolumn{1}{c}{\bf (degrees)}  & {\bf (h m s)}&{\bf ($^{o}$ $'$ $``$)}&  &&&&&&&&{\bf (degrees)}\\  \hline \hline

G\,031.061+0.094$^{3}$	& 18 47 41.342	& --01 37 21.31	&	16	&	16.2	&	15,17	&	$<$0.5	&		&--102,186	&		&	--	&	G\,31.06+0.09		 \\
G\,031.282+0.062$^{3,(7)}$	& 18 48 12.385	& --01 26 22.60	& 	83	& 110.0	&	102,113	&	15	& 110.4	& 105.1,113.0	&	25.8	&	--	&	G\,31.28+0.06		\\
G\,031.411+0.307$^{3,(7)}$	& 18 47 34.314	& --01 12 47.14	& 	9	&103.4	&	90,108	&	1.5	& 	95.7	&	94.1,95.8	&	0.6	&	y	&	G\,31.41+0.30		 \\ 
G\,033.13--0.09$^{5,(7)}$      	& 18 52 07.2	& +00 08 06	&	12.4	&	73	&	71,81	&	 0.35	&  73.3	&	72.9,73.3	&	0.1	&	--	&	G\,33.13--0.09		 \\
G\,034.246+0.134$^{\alpha,(7)}$	& 18 53 21.46	& +01 13 54	&	10.6	&	55.4	&	55,63	&	1.3	&	61.2	&	61.0,61.8	&	0.8	&	--	&	G\,34.24+0.13		 \\ 
G\,035.025+0.350$^{\alpha,(*)}$     & 18 54 00.66	& +02 01 18	&	36	&	44.3	&	40,47	& 	0.7	&	42.2	&	42.0,43.4	&	0.2	&	--	&	G\,35.02+0.35		 \\
G\,037.47--0.11$^{6,(*)}$       	& 19 00 06.7	& +03 59 27	& 	12.5	&	58.4	&	54,64	&	2.0	&	63.0	&	58.6,63.1	&	1.2	&	--	&	G\,37.475--0.106	 \\
G\,049.470--0.371$^{1,(*)}$ 	& 19 23 37.90	& +14 29 59.4	&	9.2	&	64.0	&	55,76	& 	2.0	&	57.8	&	57.0,64.0	&	1.2	&	--	&	G\,49.472--0.366	 \\
G\,049.489--0.369$^{1,(7)}$ 	& 19 23 39.83	& +14 31 05.0	&	26	&	56.1	&	55,61	& 	4.8	&	56.2	&	55.6,57.3	&	2.1	&	--	&	G\,49.49--0.37 \\\hline
\end{tabular}
\end{table*}
\clearpage

\begin{table*}
  \caption{Absorption features that we detect at 12.2-GHz towards 6.7-GHz methanol masers. The first three columns in this table are as in Table~\ref{tab:sources}. Column 4 indicates whether or not emission at 12.2-GHz was detected, followed in columns 5 to 7 by the flux density of the absorption, peak velocity (where the peak corresponds to the minimum in the absorption trough) and velocity range of the 12.2-GHz absorption features that we detect.}
  \begin{tabular}{lllrrcrrcllll} \hline
    \multicolumn{1}{c}{\bf Methanol maser} & {\bf RA}  & {\bf Dec} & {\bf 12-GHz} & {\bf Abs. Flux} & {\bf Peak} & {\bf Velocity} \\
   \multicolumn{1}{c}{\bf ($l,b$)}& {\bf (J2000)} & {\bf (J2000)}   & {\bf emission} & {\bf Density} &{\bf Velocity} & {\bf Range}\\	
      \multicolumn{1}{c}{\bf (degrees)}  & {\bf (h m s)}&{\bf ($^{o}$ $'$ $``$)}& {\bf ?}  &  {\bf (Jy)} & {\bf (\kmsns)}  & {\bf (\kmsns)}\\ \hline \hline
G\,206.543--16.355$^{2}$	& 05 41 44.15		& --01 54 44.9		&	n	&	--0.3	&	9	&	8,11		\\
G\,269.153--1.128$^{1}$		& 09 03 33.46		& --48 28 02.6		&	n	&	--0.3	&	10	&	8,12		\\
G\,301.136--0.226$^{1}$		& 12 35 35.14		& --63 02 32.6		& 	n	&	--0.2	&	--39	&	--42,--37	\\
G\,305.200+0.019$^{1}$   	& 13 11 16.93		& --62 45 55.1		& 	y	&	--0.2	&	--34	&	--40,--26	\\
G\,305.208+0.206$^{1,(7)}$	& 13 11 13.71		& --62 34 41.4		& 	y	&	--0.4	&	--40	&	--49,--30	\\
G\,318.948--0.196$^{1,(7)}$	& 15 00 55.39		& --58 58 52.8		&	y	&	--0.5	&	--37	&	--39,--27	\\
G\,323.740--0.263$^{1,(7)}$	& 15 31 45.45   	& --56 30 50.1		& 	y	&	--0.6	&	--47	&	--58,--43	\\
G\,331.278--0.188$^{1,(7)}$	& 16 11 26.59		& --51 41 56.7		& 	y	&	--0.4	&	--77	&	--91,--75	\\
G\,000.496+0.188$^{1}$		& 17 46 03.96		& --28 24 52.8		& 	y	&	--0.3	&	3	&	--14,6	\\
G\,008.669--0.356$^{1}$		& 18 06 18.99		& --21 37 32.2		& 	y	&	--0.4	&	37	&	34,50	\\
G\,008.683--0.368$^{1,(7)}$	& 18 06 23.49		& --21 37 10.2		&	y	&	--0.4	&	37	&	36,48	\\
G\,010.444--0.018$^{1,(7)}$	& 18 08 44.88		& --19 54 38.3		& 	y	&	--0.3	&	70	&	63,77	\\
G\,010.627--0.384$^{1,(7)}$	& 18 10 29.22		& --19 55 41.1		& 	y	&	--0.5	&	--1	&	--5,3		\\
G\,011.936--0.616$^{1,(7)}$	& 18 14 00.89		& --18 53 26.6		&	y	&	--0.5	&	39	&	27,46	\\
G\,012.025--0.025$^{1,(7)}$	& 18 12 01.86		& --18 31 55.7		& 	y	&	--0.3	& 	113	&	103,116	\\
G\,012.909--0.260$^{1,(7)}$	& 18 14 39.53		& --17 52 00.0		& 	y	&	--0.3	&	37	&	32,47	\\
G\,016.86--2.15$^{3}$     		& 18 29 24.411		& --15 16 04.12		& 	n	&	--0.2	&	19	&	17,21	\\
G\,023.437--0.184$^{1,(7)}$	& 18 34 39.25		& --08 31 38.5		&	y	&	--0.7	&	101	&	90,109	\\
G\,023.440--0.182$^{1,(7)}$	& 18 34 39.18		& --08 31 25.3		& 	y	&	--0.6	&	101	&	91,112	\\
G\,025.710+0.044$^{3,(8)}$	& 18 38 03.148		& --06 24 14.80		& 	y	&	--1.7	&	94	&	84,100	\\
G\,025.826--0.178$^{3}$		& 18 39 03.631		& --06 24 09.64		& 	y	&	--0.3	&	94	&	86,95	\\
G\,028.305--0.387$^{3}$		& 18 44 21.988		& --04 17 38.42		&	y	&	--0.3	&	85	&	76,87	\\
G\,029.956--0.016$^{3,(7)}$	& 18 46 03.763		& --02 39 19.69		& 	y	&	--0.4	&	100	&	90,108	\\
G\,029.979--0.047$^{3,(7)}$	& 18 46 12.974		& --02 38 58.05		& 	y	&	--0.3	&	100	&	93,108	\\
G\,030.704--0.068$^{3}$		& 18 47 36.801		& --02 00 48.99		& 	n	&	--0.4	&	91	&	89,95	\\
G\,030.818--0.057$^{3,(7)}$	& 18 47 46.984		& --01 54 19.61		&	y	&	--0.4	&	97	&	94,99	\\
G\,030.898+0.162$^{3}$		& 18 47 09.133		& --01 44 10.29		&	y	&	--0.4	&	105	&	95,106	\\
G\,031.282+0.062$^{3,(7)}$	& 18 48 12.385		& --01 26 22.60		& 	y	&	--0.9	&	109	&	100,116	\\
G\,031.411+0.307$^{3,(7)}$	& 18 47 34.314		& --01 12 47.14		& 	y	&	--0.2	&	99	&	96,101	\\ \hline

\end{tabular}\label{tab:abs}
\end{table*}

\subsection{Comments on individual sources}\label{Section:sources}

{\em G\,192.600--0.048 :} The weak 12.2-GHz methanol maser emission towards this 6.7-GHz methanol maser is one of the two weakest of our detections. The emission has a peak flux density of 0.47~Jy, which corresponds to a 4.7-$\sigma$ detection. Furthermore, the velocity of the 12.2-GHz emission is 3.6 \kmsns, close to the peak of the 6.7-GHz methanol maser at 5 \kmsns, well within the full velocity range of the associated 6.7-GHz methanol maser.

{\em G\,284.352--0.419 :} \citet{Caswell95b} detected a 0.7 Jy 12.2-GHz methanol maser towards this 6.7-GHz methanol maser in 1992. We detect no hint of emission at 12.2-GHz and expect that this is due to source variability. Due to this we have excluded this source from our statistical analysis.

{\em G\,290.374+1.661 :} Observations in 2008 June detected marginal 12.2-GHz emission toward this 6.7-GHz methanol maser outside the velocity range of the 6.7-GHz maser emission at a peak velocity of --35.1 \kms of 0.6~Jy. Follow-up observations carried out in 2008 December uncovered no detectable 12.2-GHz emission. We have therefore listed this source as having no associated 12.2-GHz emission but have removed this source from our statistical analysis to avoid the possibility that we have misclassified this source.

{\em G\,291.579--0.431 :} During the 2008 June observations we detected weak 12.2-GHz methanol maser emission towards this source. The velocity of the detected emission was within the velocity range of the 6.7-GHz methanol maser emission, however, since the detection was less than 5-$\sigma$ the source was re-observed in 2008 December and resulted in no detectable emission. As for G\,290.374+1.661, we have removed this source and the associated 1.2-mm dust clump from our statistical analysis.

{\em G\,305.208+0.206 :} Observations towards 6.7-GHz methanol maser G\,305.208+0.206 as well as near-by G\,305.202+0.208, offset by 22 arcsec, were carried out in adjacent observations. The relative amplitudes of the detected emission at these two positions suggest that all of the emission is associated with G\,305.208+0.206; however, the peak of the 12.2-GHz emission is at the peak velocity of nearby G\,305.202+0.208 and observations by \citet{Caswell95b} also indicate that there is 12.2-GHz emission associated with each 6.7-GHz methanol maser source. We suggest that either pointing errors or unfavourable weather conditions have resulted in the relative amplitudes of the emission towards the two sources being affected. Both of these methanol maser sites are located within the one dust clump.

{\em G\,305.366+0.184, G\,000.496+0.188, G\,029.865--0.043 :} These sources all have stronger 12.2-GHz methanol maser emission than the values for their 6.7-GHz counterparts as shown in Table~\ref{tab:sources}. As masers are intrinsically variable, measurements of the 6.7-GHz methanol maser flux density would need to be taken at the same epoch as the 12.2-GHz methanol masers to be certain that this was not a result of source variability.

{\em G\,034.246+0.134 and G\,035.025+0.350 :} The positions for these sources were determined with the ATCA with a large declination uncertainty of $\pm$15 arcsec (James Caswell, private communication) owing to the highly elongated beam for sources near declination 0\degrees. 

{\em G\,049.489--0.369 :} We observed this source towards a position offset from the true position by $\sim$66 arcsec (58 percent of the HPBW). We therefore expect that the flux density of this source is underestimated by approximately a factor of two. 

\begin{figure*}
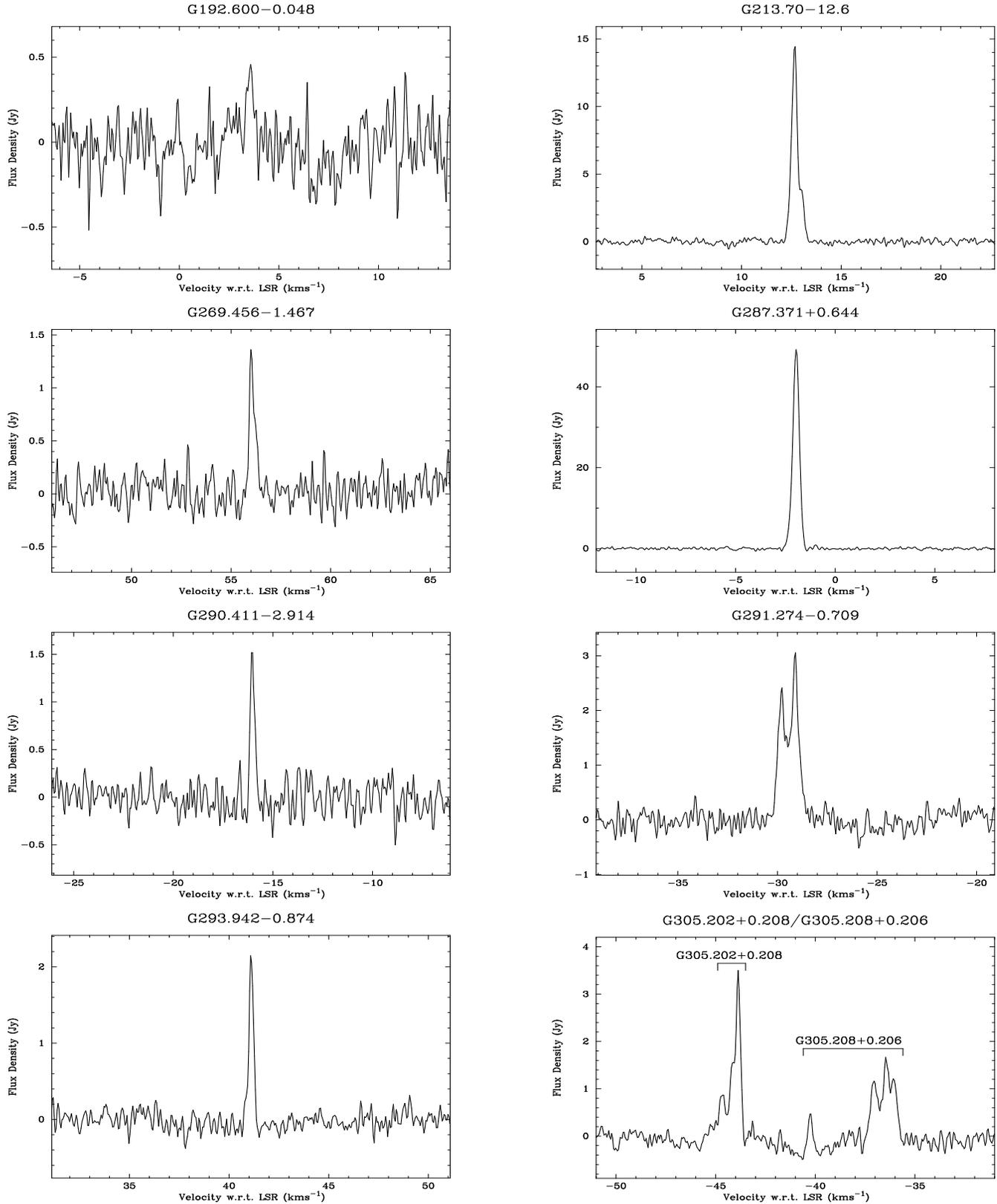

\specdfig{fig1_1}{fig1_2}
\specdfig{fig1_3}{fig1_4}
\specdfig{fig1_5}{fig1_6}
\specdfig{fig1_7}{fig1_8}
\caption{Spectra of all the 12.2-GHz methanol masers detected in the search towards 6.7-GHz methanol masers.}
 \label{fig:spectra}
 \end{figure*}
\begin{figure*}\addtocounter{figure}{-1}

\specdfig{fig1_9}{fig1_10}
\specdfig{fig1_11}{fig1_12}
\specdfig{fig1_13}{fig1_14}
\specdfig{fig1_15}{fig1_16}
\caption{-- {\emph {continued}}}
\end{figure*}
\begin{figure*}\addtocounter{figure}{-1}

\specdfig{fig1_17}{fig1_18}
\specdfig{fig1_19}{fig1_20}
\specdfig{fig1_21}{fig1_22}
\specdfig{fig1_23}{fig1_24}
\caption{-- {\emph {continued}}}
\end{figure*}
\begin{figure*}\addtocounter{figure}{-1}

\specdfig{fig1_25}{fig1_26}
\specdfig{fig1_27}{fig1_28}
\specdfig{fig1_29}{fig1_30}
\specdfig{fig1_31}{fig1_32}
\caption{-- {\emph {continued}}}
\end{figure*}
\begin{figure*}\addtocounter{figure}{-1}

\specdfig{fig1_33}{fig1_34}
\specdfig{fig1_35}{fig1_36}
\specdfig{fig1_37}{fig1_38}
\specdfig{fig1_39}{fig1_40}
\caption{-- {\emph {continued}}}
\end{figure*}
\begin{figure*}\addtocounter{figure}{-1}

\specdfig{fig1_41}{fig1_42}
\specdfig{fig1_43}{fig1_44}
\specdfig{fig1_45}{fig1_46}
\specdfig{fig1_47}{fig1_48}
\caption{-- {\emph {continued}}}
\end{figure*}
\begin{figure*}\addtocounter{figure}{-1}

\specdfig{fig1_49}{fig1_50}
\specdfig{fig1_51}{fig1_52}
\specdfig{fig1_53}{fig1_54}
\specdfig{fig1_55}{fig1_56}
\caption{-- {\emph {continued}}}
\end{figure*}
\begin{figure*}\addtocounter{figure}{-1}

\specdfig{fig1_57}{fig1_58}
\specdfig{fig1_59}{fig1_60}
\specdfig{fig1_61}{fig1_62}

\caption{-- {\emph {continued}}}
\end{figure*}
\begin{figure*}\addtocounter{figure}{-1}
\specdfig{fig1_63}{fig1_64}
\specsfig{fig1_65}
\caption{-- {\emph {continued}}}
\end{figure*}

\subsection{Statistical analysis}\label{Section:stats}

In order to determine if there are any differences between 1.2-mm dust clumps without associated 6.7-GHz methanol masers, those associated with 6.7-GHz methanol masers only and those associated with both 6.7- and 12.2-GHz methanol maser emission, we carried out a statistical analysis following the method used by \citet{Breen07} to investigate water maser presence towards $^{13}$CO clumps and 1.2-mm dust clumps. Additionally, the possibility of a relationship between dust clump properties and OH maser presence as well as peak 6.7-GHz methanol maser luminosity values and the presence of detectable radio continuum were investigated. Due to the difficulty in determining meaningful uncertainties for many of the tested source characteristics, all values have been equally weighted and assumed to be error free in all subsequent statistical analysis.

Of the 404 1.2-mm dust clump sources listed by \citet{Hill05}, 26 were excluded from our statistical analysis as they were either missing a value for one or more of the dust clump properties to be tested, or were listed by \citet{Hill05} as having mass values that are uncharacteristic of massive star formation regions. To investigate whether statistically significant relationships exist between the dust clump properties and the associated methanol masers the dust clumps were split into groups depending upon the presence or absence of specific associated phenomena.  Four different categorizations were tested, specifically:

\begin{itemize}
\item dust clumps with and without associated 6.7-GHz methanol maser emission; 
\item dust clumps with 6.7-GHz methanol maser emission with and without associated 12.2-GHz methanol maser emission; 
\item dust clumps with and without detectable radio continuum; 
\item dust clumps associated with more luminous 6.7-GHz methanol masers together with those associated with less luminous methanol maser sources (sources in this group were split into presence/absence data at the median value of 6.7-GHz peak luminosity, i.e. presence meaning the 1.2-mm dust clump is associated with a 6.7-GHz methanol maser with a peak luminosity greater than the median value and vice versa for the absence data). 
\end{itemize}
A Binomial generalized linear model (GLM) \citep{Mccul} was then fitted to the maser presence/absence data using the dust clump properties as predictors for each of the four groups. The significance of each individual clump property was tested by fitting all possible
single term models, and compared by analysis of deviance
to the null model consisting of only an intercept. Stepwise model
selection based on the Akaike Information Criteria (AIC) \citep{Burnham}
was used to select the simplest model with the greatest
predictive power and is a trade-off between goodness of fit and
model complexity. 

For each of the 1.2-mm dust clumps \citep{Hill05}, all of the derived clump properties in each of the four groups were tested: the integrated flux density (Jy), peak flux density (Jy beam$^{-1}$), source full width at half maximum (FWHM) (arcsec), distance (kpc), mass (M$_{\odot}$), radius (pc) and H$_{2}$ number density (cm$^{-3}$). In the following sections we consider p-values of less than 0.05 to be statistically significant (i.e. the hypothesis that the single term model provides no better fit than the null model consisting only of an intercept is rejected when the p-value is less than 0.05). Individual p-values, detailed tables and box plots relating to all subsequent statistical analysis are presented in Appendix~\ref{appendix}. A summary of the main results from the statistical analysis is presented in Section~\ref{sect:sum}.

\citet{Hill09} used spectral energy distribution (SED) modelling to determine a statistically probable range of values for temperature, mass and luminosity for around half of the \citet{Hill05} sample. We have not incorporated the data from \citet{Hill09} as it is only a subset of the sample of \citet{Hill05} and appears to be biased towards including a large fraction of the more evolved sources. \citet{Hill09} commented that a large fraction of the 1.2-mm dust clumps with no associated methanol masers or radio continuum do not have {\em MSX} counterparts. As a result, a number of these sources were initially included in the sample but were later excluded as the models were poorly constrained \citep{Hill09}. Furthermore, 50 of the 1.2-mm dust clumps in \citet{Hill09} are associated with 8-GHz radio continuum emission compared with 71 sources in the complete sample of \citet{Hill05}. \citet{Hill09} find that the masses of sources calculated using their SED method are similar to those presented in \citet{Hill05} with both publications showing that the ``MM-only" sources are the least massive, followed by the sources associated with 6.7-GHz methanol masers, and then the sources with radio continuum and finally the sources associated with both radio continuum and 6.7-GHz methanol maser emission. This means that even though the masses in \citet{Hill05} were calculated using a constant temperature of 20 K the calculated masses are not significantly different from the masses presented in \citet{Hill09} which were calculated using temperatures that were determined from their SED modelling.

For all dust clump properties with near/far distance ambiguities we have used the values calculated at the near distance. While it is often considered that, where distance ambiguities exist, the near distance is statistically more likely to be correct than the far distance \citep[e.g.][]{Sob05}, \citet{Xu09} suggest that using the far distances results in a better correlation between the locations of methanol masers with respect to the spiral arms of the Galaxy. In order to insure that the assumption that sources are more likely to be located at the near distance did not introduce any significant bias, the distribution of dust clump properties mass and radius of \citet{Hill05} was compared to the distribution of dust clump properties observed by \citet{Mook04}. \citet{Mook04} carried out 1.2-mm dust clump observations towards the G333 giant molecular cloud (GMC) associated with RCW106. This GMC is located at a distance of 3.6 kpc \citep{Lock79}, and all clump properties were calculated at this distance.  $^{13}$CO observations towards the G333 GMC \citep{B06} showed that, while there was emission detected along the line-of-sight, the brightest broadest feature seen in the mean velocity profile could be attributed to the target GMC. It is therefore likely that the majority of the dust clumps observed by \citet{Mook04} have been correctly assigned the distance of the GMC, introducing little bias in the clump property distributions. Comparison between the distributions of dust clump mass and radius of the dust clumps from \citet{Hill05} with the dust clumps of \citet{Mook04} has shown that the distribution of the clump properties at the near distances of \citet{Hill05} are similar to the distribution of the clump properties of \cite{Mook04}. We therefore expect that our assumption that the near distances are correct will introduce no gross bias into our statistical analysis. Peak luminosities of the 6.7-GHz methanol masers were calculated using the 6.7-GHz methanol maser peak flux density (as integrated flux densities are not readily available) and the near kinematic distance of the maser sources as used in \citet{Hill05}. 

In Sections ~\ref{Section:s6_w6} and ~\ref{Section:6_6_12} only 101 of the 1.2-mm dust clumps sources with associated 6.7-GHz methanol masers are considered. Five 1.2-mm dust clumps were removed from the observed sample of 106 dust clumps with associated 6.7-GHz methanol masers for the following reasons; G\,259.94--0.04 and G\,331.28--0.19 have some clump properties missing, the 6.7-GHz methanol maser associated with G\,284.35--0.42 had previously had a 12.2-GHz counterpart observed by \citet{Caswell95b} that we failed to detect, G\,291.58--0.53 and G\,290.374+1.661 were excluded because we detected marginal 12.2-GHz emission towards these sources in 2008 June; but not 2008 December making these sources impossible to classify and G\,6.60--0.08 because \citet{Hill05} suggest that the derived mass of this source is uncharacteristic of massive star formation regions and were therefore doubtful that it is located at the near distance.

 \subsubsection{1.2-mm dust clumps with and without 6.7-GHz methanol masers}\label{sect:6_no6}

Although the observations of ~\citet{Hill05} were primarily targeted towards sites of known 6.7-GHz methanol masers, many additional 1.2-mm dust clumps were detected in the fields of the observations and have no associated 6.7-GHz methanol maser, allowing us to determine if there is any statistically significant difference between those dust clump sources with and those without associated methanol masers. The combined sample size for this analysis is 378. Fits of the single term addition Binomial model showed an increasing probability of the presence of  6.7-GHz methanol masers associated with increasing values of dust clump integrated flux density, peak flux density, FWHM, mass and radius. This means that any one of them can give an indication of the likelihood of having an associated 6.7-GHz methanol maser. Results of the single term additions are given in Table~\ref{Tab:6_no}. 

The simplest model with the greatest predictive power contains the four dust clump properties: integrated flux density, peak flux density, FWHM and density. The estimated regression relation is

\begin{eqnarray*}
  \mbox{log}{\frac{p_{i}}{1-p_{i}}}=-2.677-0.295x_{\rm Integrated}+1.172x_{\rm Peak}\\+0.030x_{\rm FWHM}+0.00000143x_{\rm Density}
\end{eqnarray*}
where {\em x}$_{\rm Integrated}$ is the 1.2-mm dust clump integrated flux density,  {\em x}$_{\rm Peak}$ is the peak flux density, {\em x}$_{\rm FWHM}$ is the source FWHM, {\em x}$_{\rm Density}$ is the H$_2$ number density and {\em p$_{i}$} is the probability of finding a 6.7-GHz methanol maser towards the {\em i$^{th}$} 1.2-mm dust clump. The regression summary of this model is shown in Table~\ref{table:6_no_regression}.

However, the misclassification rate for this model is relatively high. Setting the threshold probability to 0.5 the model misclassifies almost two thirds of the 1.2-mm dust clumps with associated 6.7-GHz methanol masers as having no associated methanol masers and, while lowering the probability threshold allows more of the methanol maser associated 1.2-mm dust clumps to be correctly classified, it results in a marked increase in the number of 1.2-mm dust clumps with no associated methanol masers wrongly classified as having an associated methanol source. A likely explanation for this is that all of the properties involved in the regression model show a significant overlap in the range of values for the 1.2-mm dust clump with and without associated methanol maser sources (see Fig.~\ref{fig:6_no6_box}) thereby leading to the misclassification of many sources. However, it is also possible that some of the assumptions made by \citet{Hill05} in calculating the dust clump properties (like uniform temperature) could account for this. Therefore, a reliable predictive model of 6.7-GHz methanol maser presence towards 1.2-mm dust clumps requires more accurately determined dust clump properties and additional information, possibly from other wavelengths. This is in contrast to water maser associations with 1.2-mm dust clumps, for which \citet{Breen07} found a promising model that was based on only 1.2-mm dust clump radius.

\citet{Bains09} compared the locations of 6.7-GHz methanol masers \citep{Ellingsen05} with both $^{13}$CO \citep{B06} and 1.2-mm dust clumps \citep{Mook04} and used an identical statistical analysis to investigate the differences between the clumps with associated methanol masers and those without. \citet{Bains09} found that 1.2-mm dust clumps with associated methanol masers had larger radii, mass and H$_{2}$ number densities than those clumps with no associated 6.7-GHz methanol masers. The results of \citet{Bains09} seem to be in contradiction to the results that we find here: that there is no statistical difference between the 1.2-mm dust clumps H$_{2}$ number densities for sources with associated 6.7-GHz methanol masers compared to those sources with no associated 6.7-GHz methanol masers. However, this apparent discrepancy can be explained by the consideration of a few differences between the data sets. Comparison of the distribution of 1.2-mm dust clumps from \citet{Hill05} and \citet{Mook04} uncovers the first such difference. \citet{Mook04} observed 95 1.2-mm dust clumps within a giant molecular cloud (GMC) having H$_{2}$ number densities in the range 0.5 to 6 $\times$ 10$^{4}$ cm$^{-3}$. These observations had a residual noise limits of 23~mJy for the majority of the observed region and up to 50~mJy for the northernmost section of the GMC. In comparison, \citet{Hill05} observed a number of individual sources at a range of Galactic longitudes with slightly higher residual noise of 50~mJy for the majority of sources and up to $\sim$150~mJy for some sources. The H$_{2}$ number densities derived for the 404 1.2-mm dust clumps by \citet{Hill05} are all in the range 1.4 $\times$ 10$^{3}$ to 1.9 $\times$ 10$^{6}$ cm$^{-3}$, a much large range than observed by \citet{Mook04}.

\citet{Breen07} produced a three color GLIMPSE image of the GMC and showed that the central section of the GMC is dominated by bright mid-infrared emission, tracing a relatively later evolutionary stage of massive star formation. The majority of the 1.2-mm dust clumps observed by \citet{Mook04} are also aligned along this section of the GMC, meaning that the majority of the 1.2-mm dust sources will also be more evolved. \citet{Breen07} showed that the methanol masers within the region were located towards the periphery of the GMC away from the bright mid-infrared emission associated with the central axis of star formation, supporting the idea that methanol masers trace an early evolutionary stage of massive star formation. As \citet{Hill05} targeted many of their observations towards methanol masers we expect that the majority of sources observed will be at an earlier evolutionary phase than the majority of sources observed by \citet{Mook04}. 

The density of the 1.2-mm dust sources found to be associated with methanol masers in the GMC \citep{Bains09} have densities that would fall in the lower end of the range of the H$_{2}$ number densities we find associated with 6.7-GHz methanol masers. The small sample size of \citet{Mook04} (and therefore \citet{Bains09}) limits our ability to meaningfully compare the results of \citet{Bains09} with what we find here. Furthermore, if the difference in the nature of the 1.2-mm target sources of \citet{Hill05} and \citet{Mook04} as well as the differing sensitivities of these two sets of observations are considered it is not surprising that the results differ; and therefore this difference cannot be considered a contradiction.

\subsubsection{1.2-mm dust clumps and 6.7-GHz methanol maser luminosity}\label{Section:s6_w6}

Dust clumps associated with more luminous methanol masers and dust clumps associated with 6.7-GHz methanol masers with lower luminosities were compared to determine if there was any statistically significant dependance on 6.7-GHz methanol maser luminosity. The 101 1.2-mm dust clumps with appropriate data and associated 6.7-GHz methanol masers were split into two groups at the median value of 6.7-GHz methanol maser peak luminosity and the two groups were compared.  In addition to the dust clump properties \citep{Hill05} the significance of the presence of an associated OH maser was tested in regions common with the complete search for OH masers carried out by ~\citet{Caswell98} (this incorporates $\sim$70 per cent of the complete sample). Fits of the single term addition Binomial model to the 1.2-mm dust clump properties showed an increasing probability of more luminous 6.7-GHz methanol masers being associated with dust clump sources that have bigger radii, higher mass, lower densities and are located at further distances. A summary of the single term addition Binomial model can be found in Table~\ref{Tab:strong6weak6} and is represented graphically in the form of box plots in Fig.~\ref{fig:strong6weak6}. 

In order to determine if OH masers are preferentially associated with more luminous 6.7-GHz methanol masers we then fitted a Binomial model to the presence or absence of an OH maser. For this analysis only the sources that fell within the range of the complete search of \citet{Caswell98} were considered. We found that there is a greatly increased chance of finding an OH maser associated with the more luminous 6.7-GHz methanol masers, the p-value of 0.012, corresponding to a confidence level of 98.8 per cent.

\begin{figure}
\vspace{-1cm}
	\psfig{figure=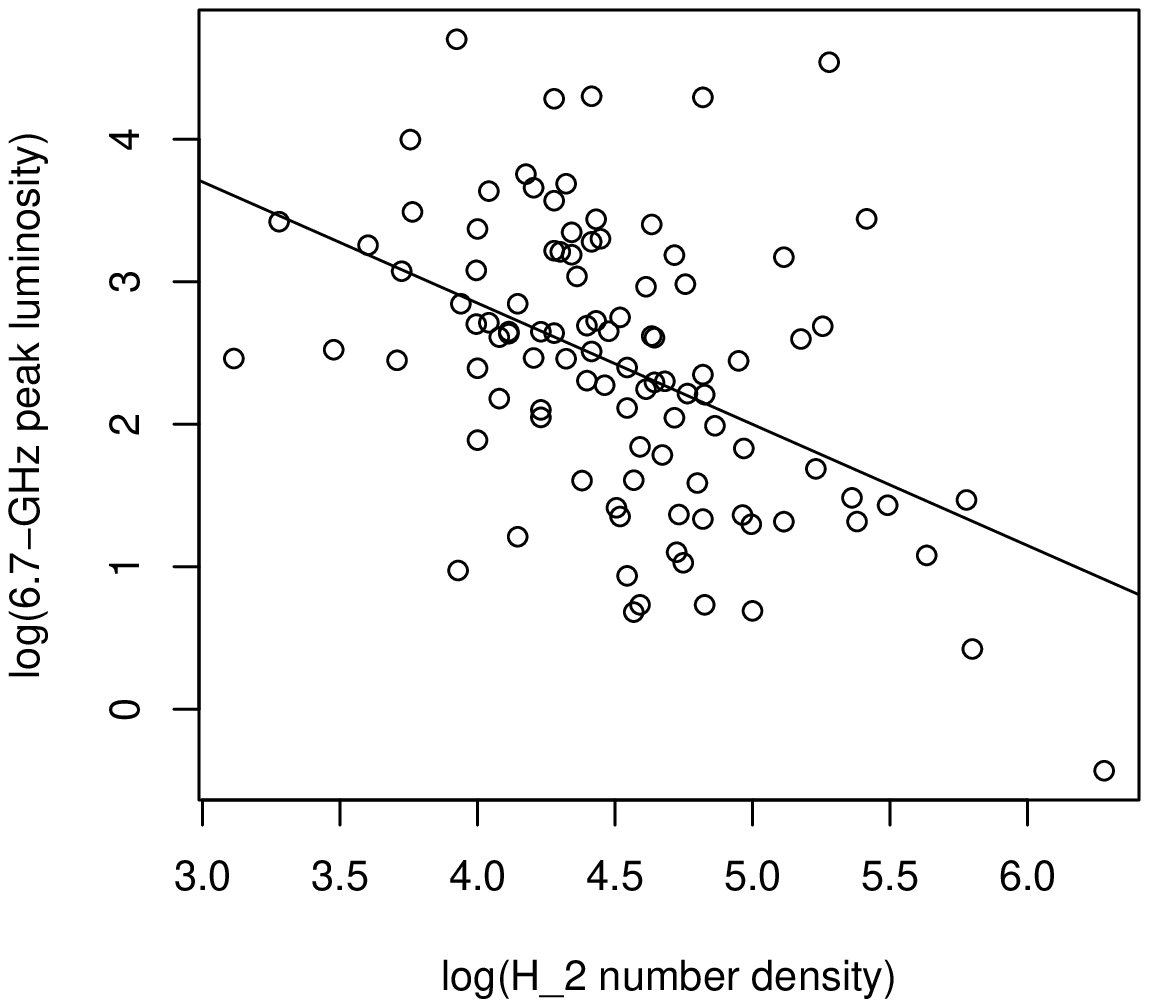,width=9cm,angle=270}
	\caption{Log of 6.7-GHz peak luminosity versus the H$_{2}$ number density of the associated dust clump in cm$^{-3}$. The line of best fit is also plotted.}
	\label{fig:6lum_density}
\end{figure}

The tendency for the more luminous 6.7-GHz methanol masers to be associated with 1.2-mm dust clumps with lower H$_{2}$ number densities was further investigated to see if there was an overall trend present in the data. Figure~\ref{fig:6lum_density} shows the log of the 6.7-GHz peak luminosity versus the log of the H$_{2}$ number density of the associated 1.2-mm dust clump. We have used the technique of linear regression and have found that there is a statistically significant negative slope in the distribution, with a weak correlation (correlation coefficient of 0.46) between the data points. The null hypothesis that there is zero slope in the data can be rejected (p-value of 1.16e-06). The line of best fit has the equation

\begin{eqnarray*}
  \mbox{log(6.7-GHz lum)}=-0.85[0.16](\mbox{log}(\mbox{H}_2 \mbox{density}))+6.25[0.75],
  \end{eqnarray*}

where `6.7-GHz lum' is the peak luminosity of the 6.7-GHz methanol maser and `H$_{2}$density' is the H$_{2}$ number density of the associated 1.2-mm dust clumps and the numbers contained in the square brackets are the standard errors of the slope and the y intercept respectively. It is likely that at least some of the scatter present in the distribution of data points can be explained by the use of the near kinematic distances which are prone to errors. A further contributer is likely to be the variable nature of the maser sources \citep[see e.g.][]{Goed}. 

The most parsimonious model produced by fitting a GLM to the high and low 6.7-GHz methanol maser luminosity associated dust clumps includes both the dust clump FWHM and the distance. The estimated regression relation is

\begin{eqnarray*}
  \mbox{log}{\frac{p_{i}}{1-p_{i}}}=-3.184+0.017x_{\rm FWHM}+0.489x_{\rm Distance},
  \end{eqnarray*}
where {\em x}$_{\rm FWHM}$ is the FWHM of the 1.2-mm dust clump and {\em x}$_{\rm Distance}$ is the distance to the source. The regression summary of this model, including an estimate of errors is shown in Table~\ref{table:s6_w6_regression}.

The accuracy of this model was tested on the data and when the threshold of probability was set to 0.5 the misclassification rates were found to be promising. Of the 51 1.2-mm dust clumps with 6.7-GHz methanol masers with high luminosities, the model correctly predicts that 38 of these 1.2-mm dust clumps will have associated methanol masers with high luminosities, but falsely predicts that the remaining 13 clumps will have associated low luminosity methanol masers. Of the 50 1.2-mm dust clumps with associated low luminosity methanol masers the model predicts 39 correctly and falsely predicts that 11 will have highly luminous associated 6.7-GHz methanol masers. It is likely that these misclassification rates can be explained by a combination of the fact that we are using 6.7-GHz peak luminosity, rather than the more appropriate value of 6.7-GHz integrated luminosity, the fact that the calculated luminosities are sensitive to errors in the adopted kinematic distances and the fact that there are a number of sources with similar 6.7-GHz peak luminosities around the median value.

\subsubsection{1.2-mm dust clumps with associated 6.7-GHz methanol masers with and without 12.2-GHz methanol masers}\label{Section:6_6_12}

The properties of the 1.2-mm dust clumps associated only with 6.7-GHz methanol masers were compared to 1.2-mm dust clumps associated with both 6.7- and 12.2-GHz methanol masers; the sample size for this analysis was 101 dust clumps (those that were excluded are shown in column 12 of Table~\ref{tab:sources} and discussed in Section~\ref{Section:s6_w6}). Fits of the single term Binomial model show that there is a statistically significant increase in the likelihood of finding a 12.2-GHz methanol maser associated with increasing 6.7-GHz methanol maser flux density, 6.7-GHz methanol maser peak luminosity and decreasing values of 1.2-mm dust clump density. The results of the single term additions are in Table~\ref{Tab:612vs6only}.  The significance of the presence of an OH maser associated with the 6.7-GHz methanol masers was tested on the sample within the search region for OH masers of \citet{Caswell98}  (this incorporates $\sim$70 per cent of the complete sample). This resulted in a p-value of 0.025, meaning that 12.2-GHz methanol masers are preferentially found towards 6.7-GHz methanol masers with associated OH masers.

The luminosity of the 12.2-GHz methanol masers were investigated to determine if individual source luminosity had any relationship with either the association with 8-GHz radio continuum or the luminosity of the associated radio continuum. We found that there is no relationship between either the presence of an associated radio continuum source or the luminosity of associated radio continuum and the luminosity of the associated 12.2-GHz methanol maser.

\begin{figure}
\vspace{-1cm}
	\psfig{figure=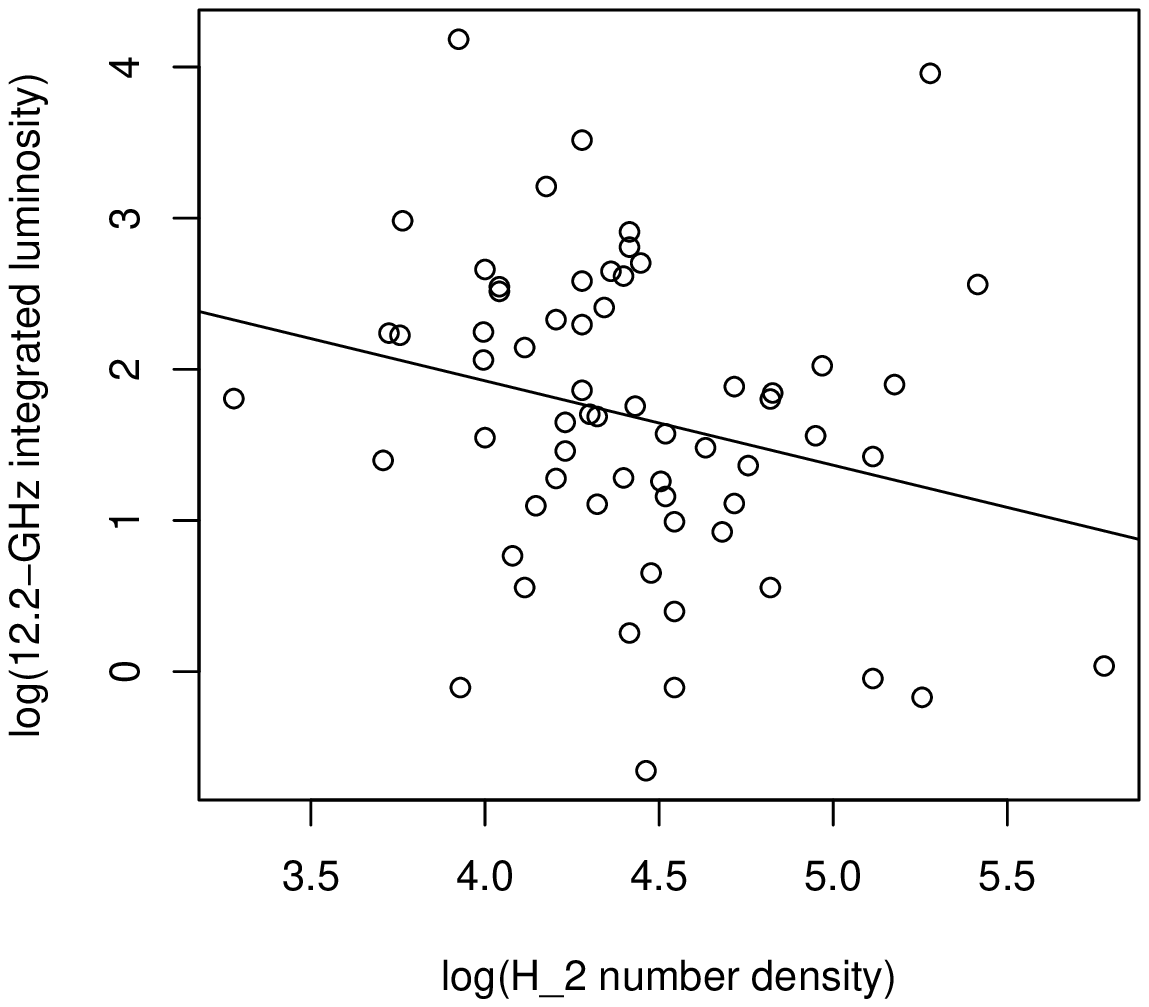,width=9cm,angle=270}
	\caption{Log of 12.2-GHz integrated luminosity versus the H$_{2}$ number density of the associated dust clump in cm$^{-3}$. The line of best fit is also plotted.}
	\label{fig:12lum_density}
\end{figure}

The luminosities of the 12.2-GHz methanol masers were compared to the H$_{2}$ number densities of the 1.2-mm dust clump using the same method as for 6.7-GHz peak luminosity in Section~\ref{Section:s6_w6}. Figure~\ref{fig:12lum_density} shows a plot of the data overlaid with the line of best fit. We find that there is a very weak correlation in this data set, with a correlation coefficient of 0.25. Similarly to Fig.~\ref{fig:6lum_density}, we expect that the largest contributions to the scatter in Fig.~\ref{fig:12lum_density} are the use of the kinematic distances and maser variability.However, there is an overall negative slope in the data that is statistically significant, the p-value is 0.05 meaning that the null hypothesis that there is zero slope can be rejected. The line of best fit has the equation 

\begin{eqnarray*}
  \mbox{log(12.2-GHz lum)}=-0.56[0.28](\mbox{log}(\mbox{H}_2 \mbox{density}))+4.16[1.23],
  \end{eqnarray*}

where `12.2-GHz lum' is the integrated luminosity of the 12.2-GHz methanol maser and the `H$_{2}$density' is the H$_{2}$ number density of the associated 1.2-mm dust clumps and the numbers contained in the square brackets are the standard errors of the slope and the y intercept respectively. 

The most parsimonious model produced by fitting a GLM to the 12.2-GHz methanol maser presence/absence data, includes only the value of the 6.7-GHz peak flux density. The estimated regression relation is

\begin{eqnarray*}
  \mbox{log}{\frac{p_{i}}{1-p_{i}}}=-0.41058+0.03571x_{\rm 6.7GHz Flux},
  \end{eqnarray*}
where {\em x}$_{\rm 6.7GHzFlux}$ is the flux density of the 6.7-GHz methanol maser peak in Jy. The regression summary of this model, including an estimate of errors is shown in Table~\ref{table:6_6_12_regression}. The physical implication of this model is that 6.7-GHz methanol masers with peak flux densities of $\sim$11.5~Jy have a probability of 0.5 of having a 12.2-GHz methanol maser counterpart with a flux density $>$0.55~Jy, with the probability increasing with increasing values of 6.7-GHz methanol maser flux density and decreasing for lower values.

For the 63 dust sources that do have associated 6.7- and 12.2-GHz methanol masers the model predicts (when the probability threshold is set to 0.5) that 44 of the 6.7-GHz methanol masers will have associated 12.2-GHz methanol maser emission and wrongly predicts that the remaining 19 will not have associated 12.2-GHz methanol maser emission. Applying the model to the 38 6.7-GHz methanol masers with no detectable 12.2-GHz emission, the model predicts 29 of these correctly and falsely expects the remaining 9 to have associated 12.2-GHz methanol maser emission.
 
 This analysis was then repeated after limiting the sample to sources where the associated 6.7-GHz methanol maser was stronger than 9~Jy (more than 16 times our detection limit of 0.55~Jy). This was done to determine if the results of the analysis of the full sample was due to a sensitivity bias that may have been introduced by the fact that 12.2-GHz methanol masers are usually weaker than their associated 6.7-GHz emission. \citet{Caswell95b} found that the median ratio of 12.2- to 6.7-GHz methanol maser intensity was 1:9 (taking into account 12.2-GHz non-detections) which we have doubled and adopted as the 6.7-GHz methanol maser cut off threshold. Therefore, methanol masers with peak flux densities of less than 9~Jy were removed from the sample, leaving 61 dust clumps, and the statistics were recalculated. The removal of the weaker 6.7-GHz methanol maser sources had little effect on the results of the statistics with the 1.2-mm dust clump density, 6.7-GHz methanol maser peak luminosity and flux density remaining the only properties showing statistically significant differences.

\subsubsection{Statistical analysis of 1.2-mm dust clumps with and without 8-GHz radio continuum}

In order to determine if there was any difference between the properties of 1.2-mm dust clumps with and without detectable radio continuum we also fitted a Binomial GLM to the radio continuum presence/absence data. 378 1.2-mm dust clumps were used in this analysis, 71 of which have detectable radio continuum \citep*[][and references therein]{Hill05}. Fits of the single term Binomial model showed that all tested 1.2-mm dust clump properties, with the exception of distance, were statistically significant. In the case of 1.2-mm integrated and peak flux densities, radius, mass and FWHM, increasing values of the dust clump properties showed a greater likelihood of being associated with detectable radio continuum, whereas in the case of the H$_{2}$ number density it is the smaller values that show an increased likelihood of associated radio continuum. Results  of the single term additions are in Table~\ref{Tab:radiocont} and that data is shown in the form of box plots in  Fig.~\ref{fig:radio}.

\subsection{Summary of statistical results}\label{sect:sum}

Here we briefly summarize the main results from the statistical analysis carried out in the previous section, Section~\ref{Section:stats}:
\begin{itemize}
	\item 1.2-mm dust clumps with associated methanol masers have higher values of integrated and peak flux densities, mass and radius than those with no associated 6.7-GHz methanol maser.
	\item 1.2-mm dust clumps with associated 6.7-GHz methanol masers with high peak luminosities are more likely to be associated with OH masers and have larger values of mass and radius and lower values of density than those clumps that are associated with 6.7-GHz methanol masers with low luminosities. 
	\item 1.2-mm dust clumps associated with both 6.7- and 12.2-GHz methanol masers are more likely to be associated with OH masers and have lower densities than 1.2-mm dust clumps with just 6.7-GHz methanol maser associations. 
	\item The luminosity of 12.2-GHz methanol masers is not dependent on either the presence of associated radio continuum, or the luminosity of associated radio continuum.
	\item 6.7-GHz methanol masers with 12.2-GHz methanol maser counterparts have larger flux densities and higher peak luminosities than 6.7-GHz methanol masers with no 12.2-GHz counterpart and this is not likely to be a result of a sensitivity bias. 
	\item 1.2-mm dust clumps with associated radio continuum have higher values of mass, radius, peak and integrated flux density and lower values of density than clumps without detectable radio continuum. 
\end{itemize}

\section{Discussion}

\subsection{Comparison between 6.7- and 12.2-GHz methanol masers}

We have made a preliminary investigation of the relative intensities and velocity ranges of the 6.7- and 12.2-GHz methanol maser sources within this sample. Detailed analysis of these properties, as well as spectral structure, will be carried out in a subsequent paper which will present 12.2-GHz observations towards all 6.7-GHz methanol masers detected in the southern hemisphere component of the Parkes Methanol Multibeam (MMB) Survey \citep{Green08}. The MMB survey for 6.7-GHz methanol masers is the most sensitive survey yet undertaken for young high-mass stars in the Galaxy and is complete within 2 degrees of the Galactic plane.

Investigation of the peak flux densities of the 6.7- and 12.2-GHz methanol maser associations reveals that there are three instances where the flux density of the 12.2-GHz methanol maser is stronger than the reported values of the associated 6.7-GHz methanol maser emission. Given that interstellar masers are intrinsically variable it is likely that in at least some of these instances the 6.7-GHz methanol masers are now stronger than their reported values from several years prior to the 12.2-GHz methanol maser observations. 

We have made a comparison of the 6.7- and 12.2-GHz methanol maser peak flux densities (where emission was detected at both frequencies) and find that the median ratio of 12.2- to 6.7-GHz methanol flux density is 1:5.9 with an average ratio of 1:3.1. We additionally find that the most extreme ratios of 12.2- to 6.7-GHz methanol maser peak flux density are 1:180 and 1:0.56, highlighting the fact that when the 12.2-GHz emission is stronger it is only marginally stronger. In comparison, \citet{Caswell95b} found a median ration of 12.2- to 6.7-GHz methanol maser peak flux density of 1:5.4.  

We have compared the ratio of 6.7- to 12.2-GHz peak flux density to both the peak luminosity of the 6.7-GHz methanol maser and the integrated luminosity of the 12.2-GHz methanol masers. We find that there is a general trend whereby the 6.7-GHz methanol masers with high peak luminosities have higher ratios of 6.7- to 12.2-GHz methanol maser flux density (i.e. the more luminous 6.7-GHz methanol masers have relatively weaker emission at 12.2-GHz). Comparison between source flux density ratios and 12.2-GHz methanol maser luminosities calculated from the integrated flux densities of the 12.2-GHz methanol maser sources show the complementary trend (i.e. the less luminous 12.2-GHz methanol masers have relatively stronger emission at 6.7-GHz). 

\begin{figure}
\vspace{-1cm}
	\psfig{figure=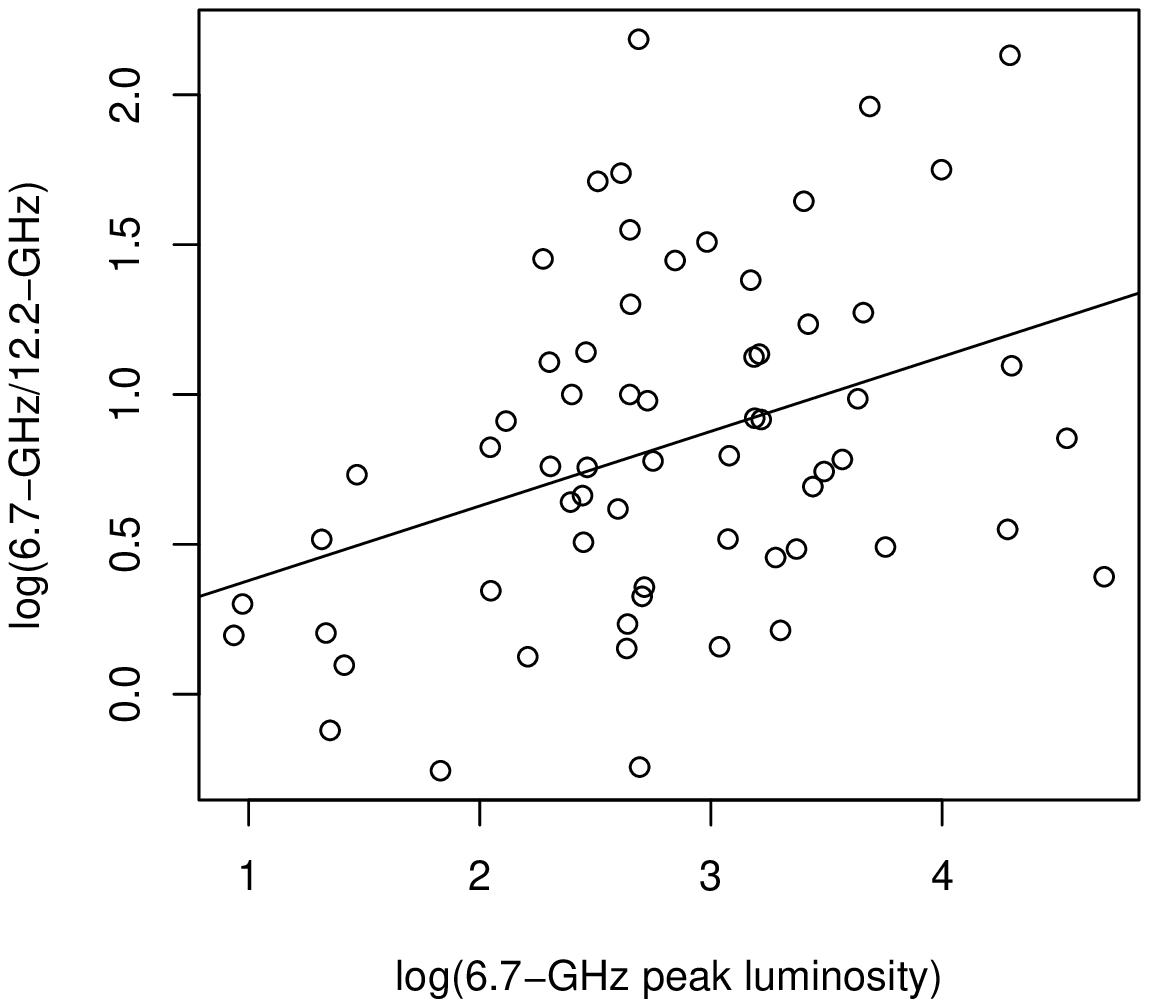,width=9cm,angle=270}
	\caption{Log of the 6.7-GHz methanol maser peak flux density divided by the 12.2-GHz methanol maser peak flux density versus the log of the 6.7-GHz peak luminosity, overlaid with the line of best fit.}
	\label{fig:6lum_v_ratio}
\end{figure}

\begin{figure}
\vspace{-1cm}
	\psfig{figure=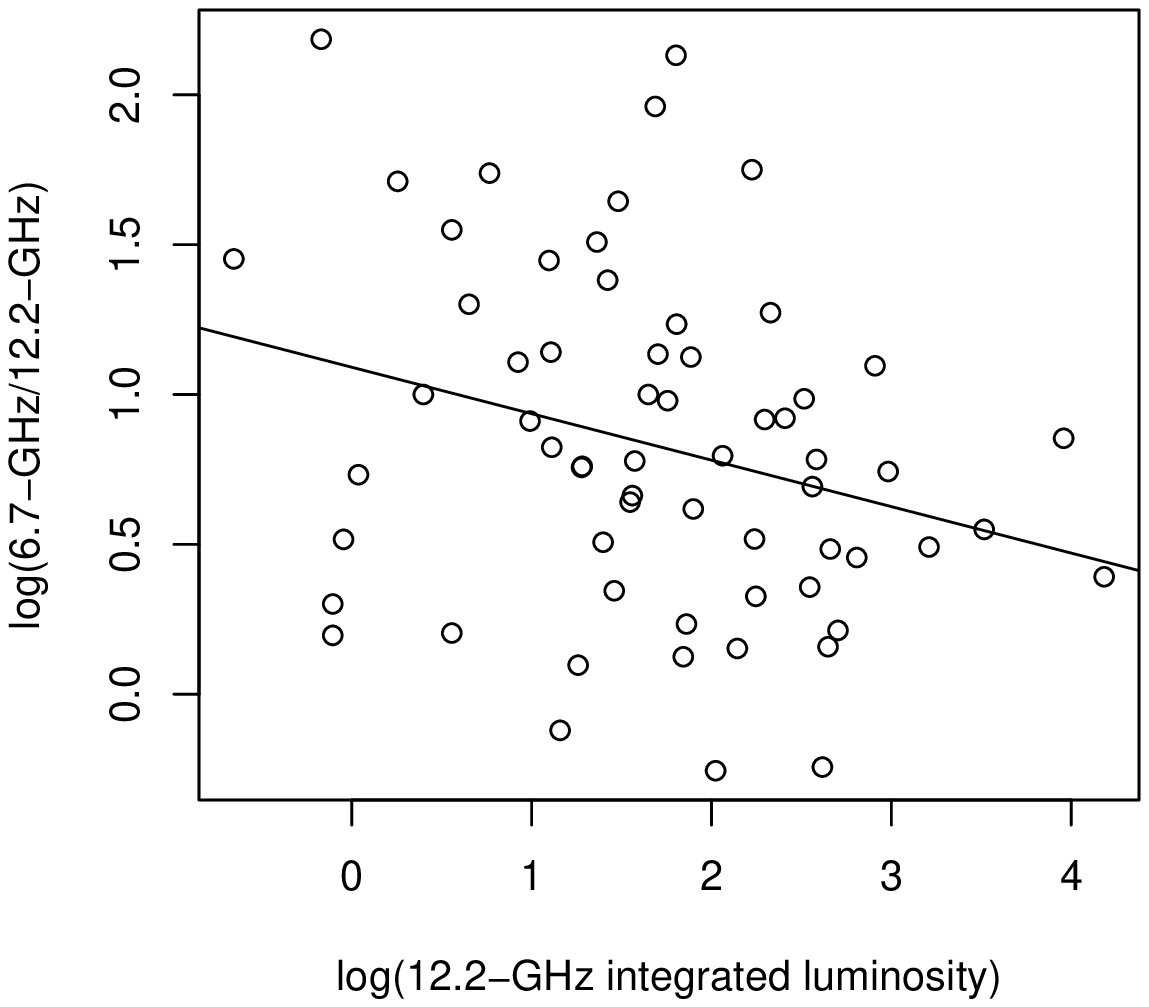,width=9cm,angle=270}
	\caption{Log of the 6.7-GHz methanol maser peak flux density divided by the 12.2-GHz methanol maser peak flux density versus the log of the 12.2-GHz integrated luminosity, overlaid with the line of best fit.}
	\label{fig:12lum_v_ratio}
\end{figure}

Plots of the flux density ratios versus 6.7-GHz peak luminosity and 12.2-GHz integrated luminosity are shown in Figs~\ref{fig:6lum_v_ratio} and \ref{fig:12lum_v_ratio} to demonstrate our results quantitatively. We have used linear regression to show that the trends in the data are statistically significant. For the 6.7-GHz peak luminosity, we find a slope of 0.25 with a standard error of 0.08 and a p-value of 0.003 which allows us to reject the null hypothesis that there is zero slope. The correlation coefficient between the points is 0.36 which is classified as a weak correlation. For the 12.2-GHz integrated luminosity we find a slope of --0.15 with a standard error of 0.07 and a p-value of 0.03 which similarly means that the slope of the line is statistically significant and non-zero. The correlation coefficient is 0.27 which also corresponds to a weak correlation between the data points. The significant scatter present in both data sets are likely to be due to a combination of a few factors. The first is that the calculated luminosities are heavily dependent on the derived kinematic distances for the sources, which in many instances have large ambiguities. We have assumed the near distances for sources here as for all previous analysis which, for some sources, may introduce large errors into the calculated source luminosities. It is also likely that the 6.7-GHz methanol maser flux densities reported in the literature would not correspond to the true flux density of these sources if measured at the time of the 12.2-GHz methanol maser observations. We therefore expect that source variability also contributes to the scatter in Figs~\ref{fig:6lum_v_ratio} and \ref{fig:12lum_v_ratio}. It is expected that these influences will have an effect on our data; however, we would expect that they would act to weaken trends not acting in a systematic way to strengthen them. It is therefore likely that accurate distance measurements and single epoch observations of both methanol maser transitions would result in a more tightly correlated relationship, but, this is something for future investigation.

In section 3.2.2 we showed that the 6.7-GHz peak luminosity increases with decreasing dust clump density.    We also show that centimetre radio continuum emission is preferentially found towards lower density dust clumps, which suggests that the dust clump density decreases as the source evolves (see section 4.3 for a more detailed discussion of our results and their implications for an evolutionary sequence for high-mass star formation).  These results (and others) combined suggest that the 6.7-GHz methanol peak luminosity increases as the high-mass star formation it is associated with evolves.  That the average ratio of 6.7- to 12.2-GHz peak flux density increases as the 6.7-GHz peak luminosity increases implies that the rate of increase of the 12.2-GHz peak flux density as the source evolves is less than that for the 6.7-GHz transition.   Modeling of methanol masers by Cragg et al. (2005) shows that the 6.7-GHz transition is essentially always stronger than the 12.2-GHz transition and our results suggest that as the underlying star formation region evolves  the relative intensity of the 6.7-GHz transition increases more quickly.  In particular, Figures 2 \& 3 of Cragg et al. (2005) show that the intensity of the 12.2-GHz transition decreases more rapidly with increasing gas temperature than the 6.7-GHz transition does, which is consistent with an evolutionary trend for the 6.7- to 12.2-GHz flux density ratio to increase.

Inspection of the velocity ranges of the 6.7-GHz methanol masers as well as their associated 12.2-GHz methanol maser emission has shown that of the 68 12.2-GHz methanol masers that we observed, 49 sources show the peak of their emission at the same velocity as the 6.7-GHz methanol maser, to within the spectral resolutions of the respective observations. For the remaining 16 sources we find the 12.2-GHz peak velocity within the velocity range of the detectable 6.7-GHz methanol maser emission and is coincident with a secondary 6.7-GHz methanol maser feature.  In the majority of cases we find that the 12.2-GHz methanol masers have narrower velocity ranges than the 6.7-GHz methanol maser counterparts. The average velocity range of the 6.7-GHz methanol masers with associated 12.2-GHz methanol masers is 9 \kms and the average velocity range of the 12.2-GHz methanol masers that we observe is 3.4 \kmsns. The median value of 12.2-GHz velocity range of the sources that we observed is 5.4 \kms which is remarkably similar to the median value of the 131 sources reported by \citet{Caswell95b} as 5 \kmsns.

\subsection{Absorption at 12.2-GHz}
29 of our target 6.7-GHz methanol masers show absorption at 12.2-GHz, of which 24 also show 12.2-GHz maser emission. In comparison \citet{Caswell95b} detected 19 absorption spectra toward 238 6.7-GHz methanol masers. We suspect that absorption accompanying emission in some cases may be an artifact of out processing method (see Section~\ref{section:obs}). 

\subsection{Evolutionary sequence for sources}

The relatively large number of class~II methanol maser sources in our Galaxy (esimated to be of order 1000), means that they must be associated with objects covering a range of stellar masses and also have an appreciable ``lifetime''.  It is generally assumed that the properties of the masers and the complementary data at different wavelengths will depend upon both the age of the source and its mass.  At present we do not have sufficient information to disentangle these two factors.  Circumstantial evidence can sometimes be used to infer one or the other as the primary factor. However, some of the attributes ascribed to evolutionary factors may be partially, or even largely mass related (and vice versa).

\citet{Purcell09} propose that the difference between radio-loud and radio-quiet massive young stellar objects (YSOs) may be attributed to evolution, with the radio-loud sources corresponding to a later stage of evolution. Our statistical comparison between 1.2-mm dust clumps with associated radio continuum from ultra-compact (UC) \ionhy regions at 8 GHz and those without has shown that 1.2-mm dust clumps with detectable radio continuum in general have bigger radii, masses, flux densities as well as lower H$_2$ number density. We therefore propose that these 1.2-mm dust clump properties give an indication of comparative stages of evolution (i.e. sources with larger radii, masses, flux densities and smaller H$_{2}$ number densities are more likely to be at a later stage of evolution than sources which have smaller values of radii, mass, flux density and larger values of density). \citet{Purcell09} similarly found that YSOs with associated 8-GHz radio continuum are more massive than those without detectable radio continuum. 

The evolutionary interpretation of the results of the statistics is contingent on the assumption that a large number of the 1.2-mm dust clump sources with no detectable radio continuum will go on to form massive stars. While it is likely that some of the 1.2-mm dust clumps without associated radio continuum are `failed' cores and will not form massive stars, we know that a large number of them will do so, since they exhibit other tracers that are indicative of massive star formation such as methanol masers \citep{Caswell09,Minier03,Walsh98,Walsh97,Caswell95a,Szy02} and ammonia \citep{Longmore07}. \citet{Hill09} concluded from investigation of a sub-set of the \citet{Hill05} sample that the 1.2-mm dust clumps with no overt signs of massive star formation displayed similar characteristics to those 1.2-mm dust clumps associated with known massive star formation regions. Therefore, as the number of sources in this sample is large we believe that the possibility of the inclusion of some `failed' cores is inevitable, but the number will be few and so will have little effect on our statistics. 

\citet{Longmore07} detected 24-GHz continuum emission towards some of the 1.2-mm dust clump sources where no 8-GHz continuum emission was detected by \citet{Walsh98}. It is therefore likely that some of the sources that we have classified as having no associated continuum emission are in fact associated with radio continuum from hyper-compact (HC) H{\sc ii} regions which are often optically thick at centimetre wavelengths and are associated with an earlier evolutionary phase of massive star formation than \UCHII regions. According to \citet{Longmore07}, this means that sources with detectable radio continuum at 24-GHz and not 8-GHz are part of another class of object that is more evolved than sources with no detectable radio continuum and less evolved than sources with detectable continuum at centimetre wavelengths. This distinction also means that we have not introduced any bias into our analysis by not accounting for sources that are in the HCH{\sc ii} region stage.

Interpretation of the relative evolution of sources associated with the different maser properties can be made by comparing the 1.2-mm dust clump properties associated with these sources to the results of the analysis with \UCHII regions. We find that 1.2-mm dust clumps with associated 6.7-GHz methanol masers have higher flux densities, radii and mass than those with no associated 6.7-GHz methanol maser. This suggests that 6.7-GHz methanol masers are present prior to any significant change in source density that occurs along side the formation of  \UCHII regions.  However, the 1.2-mm dust clump observations of \citet{Hill05} have a relatively low resolution (24\arcsec) and so only reflect the large scale structures.  Hence significant changes in source structure and physical conditions are likely to be occurring on smaller spatial and temporal scales, but these will only be revealed by higher resolution observations of the dust properties.

Our analysis of 6.7-GHz methanol maser peak luminosity and 6.7-GHz methanol masers associated with 1.2-mm dust clumps shows that methanol masers with high peak luminosities are more likely to be associated with 1.2-mm dust clumps of larger mass and clump radii and lower H$_{2}$ number densities than 6.7-GHz methanol masers with lower peak luminosities. While we find that there is no statistical difference between the flux densities of 1.2-mm dust clump sources with varying levels of 6.7-GHz methanol maser peak luminosity, we find that the properties of the 1.2-mm dust clumps associated with 6.7-GHz methanol masers with high luminosities are similar to those 1.2-mm dust clumps that are associated with a later stage in evolution. We additionally find that more luminous 6.7-GHz methanol masers are more likely to be associated with OH masers which are known to trace a generally later evolutionary phase of massive star formation than isolated methanol masers \citep[e.g.][]{FC89,Cas97}. Therefore we expect that there is a trend between 6.7-GHz methanol maser peak luminosity and evolutionary phase, with the more luminous 6.7-GHz methanol maser peaks tending towards a later phase in evolution than 6.7-GHz methanol masers with low peak luminosities. This important result will deserve further investigation using luminosities that have been calculated from integrated maser flux densities.

We find that 1.2-mm dust clump sources associated with both 6.7- and 12.2-GHz methanol masers are more likely to have lower densities than the sources which are associated with only the 6.7-GHz maser transition of methanol. Pumping models of methanol
masers \citep[e.g.][]{Cragg05} predict that the 6.7- and
12.2-GHz transitions are inverted over very similar ranges of parameter
space.  The absence of 12.2-GHz masers associated with a large fraction
of 6.7-GHz sources (40 \%) implies that the physical conditions are commonly
close to the point where the 12.2-GHz masers switch on or off.  \citet{Cragg05} also find that 6.7-GHz methanol masers can exist at higher densities than 12.2-GHz methanol masers, consistent with the results of our analysis. We also find that 12.2-GHz methanol masers are more likely to be associated with 6.7-GHz methanol masers with higher flux densities and peak luminosities. This result is not surprising as all previous observations of both 12.2- and 6.7-GHz methanol masers have shown that the 12.2-GHz emission is rarely brighter than the associated 6.7-GHz methanol maser emission. However, when we lessened the chance of a sensitivity bias by removing all of the weak 6.7-GHz methanol masers from our analysis, we find that our results do not change. This means that there is likely to be a delay between when the 6.7-GHz methanol maser `switches on' and when the 12.2-GHz methanol maser `switches on'. Furthermore, 12.2-GHz methanol masers are more likely to be found toward 6.7-GHz methanol masers with associated OH masers. We therefore suggest the 12.2-GHz methanol masers are present sometime after the onset of 6.7-GHz methanol maser emission. 

\citet{vander} estimates the lifetime of 6.7-GHz methanol masers to be in the range 2.5 $\times$ 10$^{4}$ to 4.5 $\times$ 10$^{4}$ years, combining this with our detection rate of associated 12.2-GHz methanol masers we expect that the corresponding lifetime of 12.2-GHz methanol masers will be in the range 1.5 $\times$ 10$^{4}$ to 2.7 $\times$ 10$^{4}$ years. This inference assumes that 12.2-GHz methanol masers are never without associated 6.7-GHz methanol maser emission. However, given that previous searches have not resulted in serendipitous detections of 12.2-GHz methanol masers without accompanying 6.7-GHz methanol maser emission and the fact that the current search and previous searches \citep{Caswell95b,Blas04} have shown that 12.2-GHz methanol maser emission is only rarely brighter that the associated 6.7-GHz maser emission, this is not an unreasonable assumption.

While our simplified evolutionary results appear to be self consistent and reliable the full picture is likely to be more complex.  We found that 1.2-mm dust clumps with detectable radio continuum have larger radii, masses, flux densities and smaller H$_{2}$ number densities and suggest that these properties can be used to give an indication of the probable evolutionary stage. However, it is likely that some of the more massive dust clumps will go on to form higher mass stars and are therefore more likely to have detectable radio continuum earlier than some of the less massive cores. Furthermore the flux density of the radio continuum will depend on the dust fraction and the type of star that the 1.2-mm dust clump is associated with.

The H$_{2}$ number densities of the 1.2-mm dust clumps are central to a number of our evolutionary arguments. There are several factors that directly affect the accurate determination of this quantity. Both the 1.2-mm dust clump mass and radius rely heavily on the accuracy of the distance measurements. However, it would be unlikely that the susceptibility of the radius and mass to errors would have a systematic effect that resulted in the strengthening of a relationship between radius and some other property, such as maser presence - it would be much more likely to dilute such a relationship. We therefore believe that the only effect that potential errors in the determined 1.2-mm dust clump radius (and therefore the H$_{2}$ number density) would have on our statistics is to weaken true relationships rather than falsely cause correlations between properties. 

\citet{Hill05} adopted a constant temperate of 20~K to calculate the masses of the 1.2-mm dust clumps. \citet{Hill09} have shown that in the vast majority of cases this is an overestimate but that the distribution of masses is similar to that found in \citet{Hill05}. As we expect less evolved 1.2-mm dust clumps to be colder than the more evolved sources that show detectable 8-GHz radio continuum emission, errors introduced by the assumption of a constant temperature would result in the less evolved sources having underestimated masses relative to the more evolved sources. Therefore, the use of accurate temperature measurements would strengthen our finding that more evolved sources have lower H$_{2}$ number densities. Due to this dependance however, we are unable to separate the effects that the H$_{2}$ number density and temperature have on the evolutionary phase and expect that temperature would also play a large role in indicating the relative evolutionary phase of the sources.

As \citet{Hill05} targeted their observations towards sources suspected of undergoing massive star formation, there is a possibility that an unbiased search would result in a distribution of source properties that differs from this sample. However, as discussed in Section~\ref{sect:6_no6} the distribution range of the 1.2-mm dust clump properties from the targeted sample of \citet{Hill05} greatly exceeds the distribution range from the unbiased search of a GMC \citep{Mook04}. Furthermore, as the number of serendipitous detections of 1.2-mm dust clumps within the fields of targeted objects is large (253), the likelihood of biases being introduced by the targeted nature of this sample is low. 

Another point that needs to be considered is that a dust clump is unlikely to form only one star. Some of the confusion in our predictive models can be explained by the fact that often, there are clusters of stars forming within the one dust clump, at different evolutionary stages and with different mass ranges. This fact is evident in our sample because there are multiple instances where there is more than one 6.7-GHz methanol maser within the one dust clump.

Combining all of our statistical analyses we find that as the 1.2-mm dust clump becomes more massive (n.b. this is mostly likely a consequence of the relatively underestimated masses of the less evolved sources, a consequence of the constant temperature assumption, rather than a real increase in mass as discussed above),  and increases in radius and flux density, a weak 6.7-GHz methanol maser forms. Then, as the clump becomes less dense and continues to increase in mass, radius and flux density, the 6.7-GHz methanol maser increases in flux density and the 12.2-GHz methanol maser forms. At this stage the detection of associated OH masers and 8-GHz radio continuum becomes more likely. We therefore propose the following scenario for evolutionary sequence of sources from less evolved to more evolved: (i) 1.2-mm dust clumps with no associated methanol maser emission or 8-GHz radio continuum, (ii) 1.2-mm dust clumps with associated 6.7-GHz methanol masers, (iii) 1.2-mm dust clumps with associated methanol maser emission at both 6.7- and 12.2-GHz. 8-GHz radio continuum is observed toward 29 of our sample of 113 methanol masers and we find that 19 of these also have associated 12.2-GHz methanol maser emission (this is not statistically significant p-value=0.24 from a chi-sqared test). We therefore postulate that 8-GHz radio continuum generally becomes detectable toward the end of the 6.7-GHz methanol maser only stage, prior to the onset of 12.2-GHz methanol maser emission if the mass of the star is sufficiently large. 

\begin{figure*}
	\psfig{figure=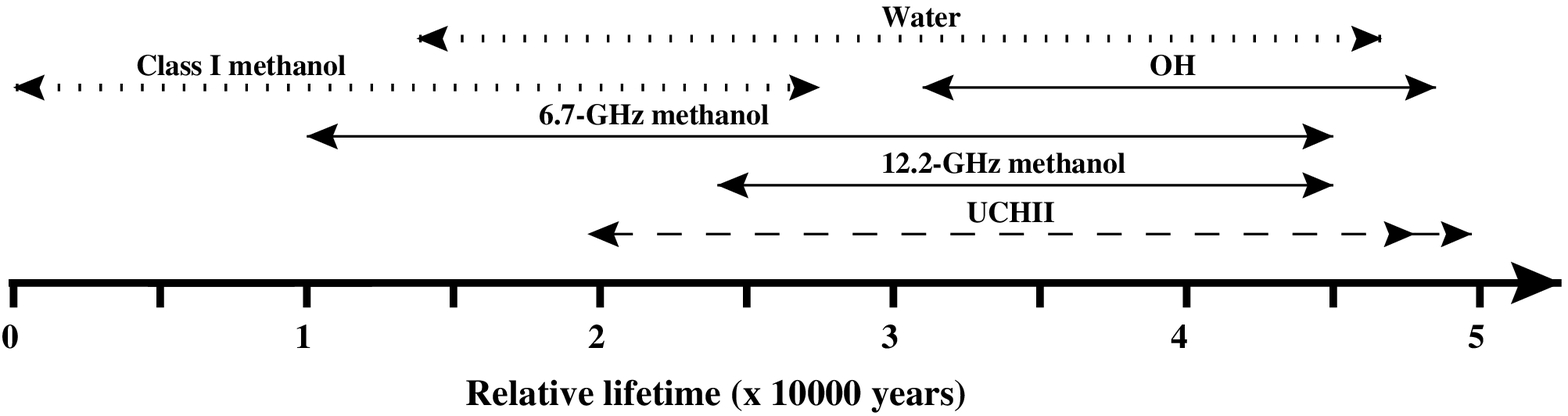,width=15cm,angle=0}
	\caption{Evolutionary sequence for masers associated with massive star formation regions. Lifetimes represented by dotted lines indicate that they have been estimated from the literature along with unpublished data. The lifetime of the \UCHII region is represented by a dashed line as the onset of radio continuum emission is particularly dependent on the mass of the associated star. The double arrow head on the far end of the \UCHII region lifetime is to show that it will persist past the end of the range of the plot. }
  \label{fig:evolution}
\end{figure*}

In Fig.~\ref{fig:evolution} we present our proposed evolutionary sequence for masers in massive star formation regions. We have used the evolutionary sequence presented in \citet{Ellingsen07} as a starting point, adjusted it according to our new data, and placed quantitative estimates on the relative lifetimes of the different species using additional maser observations \citep[e.g.][]{Caswell09}. The estimates of the lifetimes of the different maser species have been extrapolated from the 6.7-GHz methanol maser lifetime of \citet{vander} and detection statistics for the different maser species.

\subsubsection{6.7- \& 12.2-GHz methanol maser association with radio continuum}

There are 29 1.2-mm dust clumps that are associated with both methanol maser emission and 8-GHz radio continuum within our sample. 19 of these sources are associated with both 6.7- and 12.2-GHz methanol maser emission while the remaining 10 are associated with only 6.7-GHz methanol masers. We have compared these two groups of sources in order to determine if there are any differences in the 1.2-mm dust clump properties. We find that the clumps with both 12.2 and 6.7-GHz methanol masers have large FWHM (p-value 0.01), lower densities (p-value 0.07) and higher 6.7-GHz methanol maser peak luminosity and peak flux densities. We have also investigated the 8-GHz radio continuum luminosities (calculated from the flux densities given in \citet{Walsh98}) and found that there is no difference between either the flux density or the luminosity of the continuum between those sources with associated 12.2-GHz methanol masers and those without. This implies that the presence of an associated 12.2-GHz maser is much more reliant on suitable densities rather than either the presence or strength of an associated radio continuum source. In general, however, 1.2-mm dust clumps with associated radio continuum emission are more likely to satisfy the density ranges for emission from the 12.2-GHz methanol maser transition.

\subsection{Association with GLIMPSE point sources}

Positions of the 6.7-GHz methanol masers were compared to the positions of infrared point sources listed in the GLIMPSE point source catalogue as well as the GLIMPSE archive. As most methanol maser positions are known to a high precision we assume an association between the GLIMPSE sources and the methanol masers if they lie within 2 arcsec of each other. We find that 90 of the methanol masers that we observe lie within the regions covered by GLIMPSE of which 51 (57 \%) are associated with either a GLIMPSE point source or GLIMPSE archive source. \citet{Ellingsen06} found 68 \% of a sample of 56 methanol masers had associated GLIMPSE sources. The difference in detection rates is not statistically significant.

\begin{figure}
	\psfig{figure=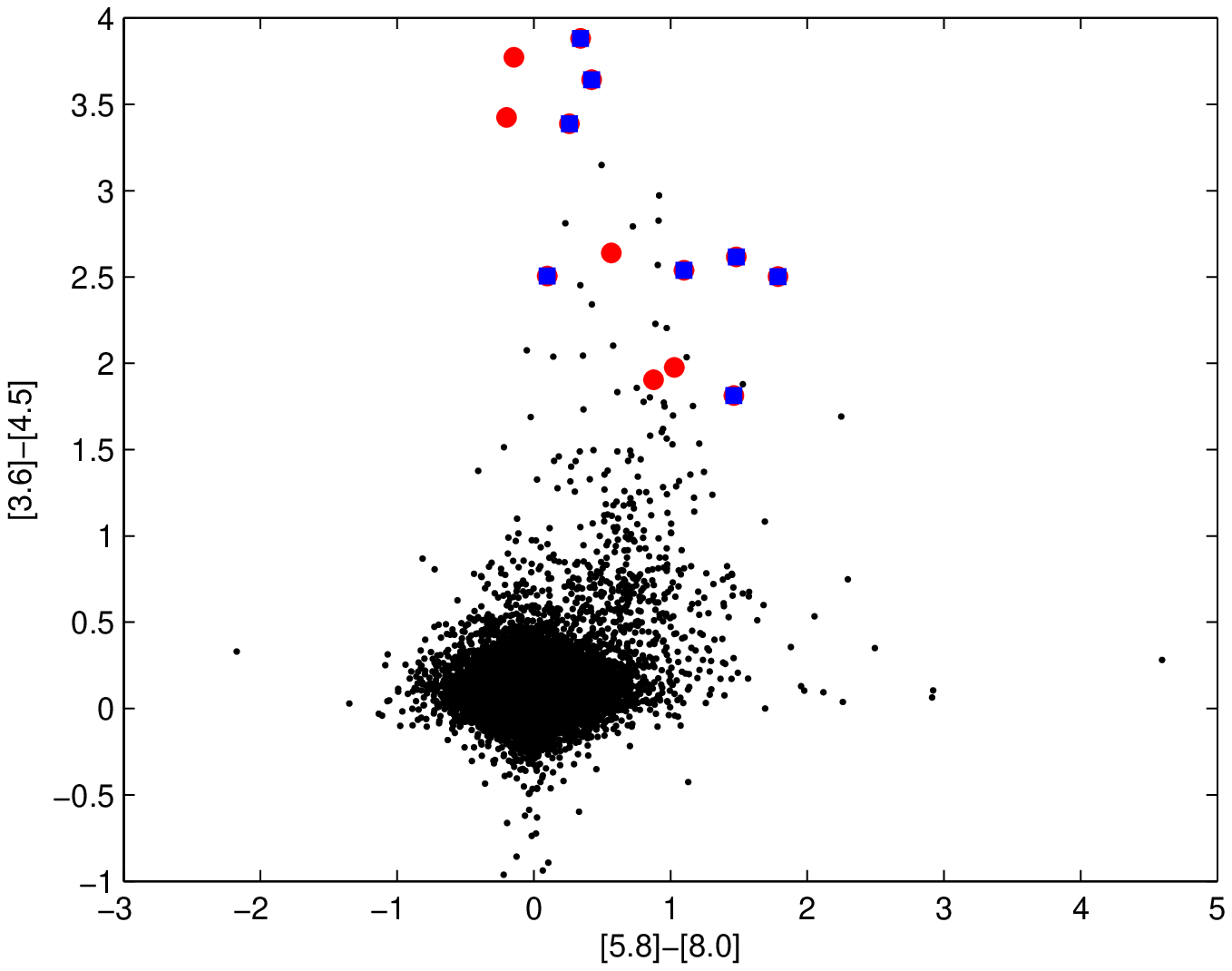,width=9cm,height=8cm,angle=0}
	\caption{Colour-colour plot of GLIMPSE point source data. Methanol maser sources showing both 6.7- and 12.2-GHz emission are represented by blue squares and sources with only the 6.7-GHz methanol maser transition are represented by the red circles. The black dots represent all of the GLIMPSE point sources within 30 arcmin of {\em l} = 326$\fdg$5, {\em b} = 0\fdg0.}
	\label{fig:glimpse}
\end{figure}

Fig.~\ref{fig:glimpse} shows a plot of the [3.6]--[4.5]~$\mu$m versus [5.8]--[8.0]~$\mu$m colours of the GLIMPSE point sources associated with the methanol masers in our sample. For sources to be present on this plot they needed to have flux density measurements for all four of the IRAC bands, limiting the number of 6.7-GHz methanol maser associated GLIMPSE sources in the plot to 13, 8 of which are also associated with 12.2-GHz methanol maser emission. Similarly to \citet{Ellingsen06}, we find that the GLIMPSE sources associated with the methanol masers lie above the majority of the comparison sources in our colour-colour plot. 

We have compared the colours of the GLIMPSE sources associated with both 6.7- and 12.2-GHz methanol masers with those associated with only 6.7-GHz methanol masers and find that there is no statistically significant difference between them. We will not be able to determine if this is a consequence of our small sample size or rather a real effect, until we can compare the colours for a larger number of sources. However, if our results are representative then this implies that the masers themselves are much more sensitive to evolutionary changes than is the associated infrared emission of wavelengths $<$ 10~$\mu$m.

\section{Conclusion}

We have searched 113 known 6.7-GHz methanol masers for associated 12.2-GHz methanol maser emission with the ATNF Parkes 64-m radio telescope. We have detected 12.2-GHz methanol masers towards 68 of these sources; a detection rate of 60 percent. We estimate the life-time of 12.2-GHz methanol maser emission to be of the order of 2.1 $\times$ 10$^{4}$ years. We find that for the majority of sources, the peak velocity of the 12.2-GHz methanol maser emission is the same as the 6.7-GHz methanol maser and, for the remainder, the 12.2-GHz peak emission is coincident with a secondary feature present in the 6.7-GHz methanol maser spectrum. Comparison between the peak flux densities of the respective methanol maser transitions has shown that median ratio of 6.7- to 12.2-GHz peak flux density is 1:5.9, similar to previous searches. We additionally find that the ratio of 6.7- to 12.2-GHz methanol maser peak flux density has some dependence on 6.7-GHz methanol maser luminosity, with the more luminous 6.7-GHz sources having comparatively weaker 12.2-GHz methanol maser emission and vice versa. 

All of the target sources have previously been observed at 1.2-mm for dust continuum emission \citep{Hill05}. We have carried out statistical analysis of the dust clump properties of sources in the following four subgroups compared the 1.2-mm dust clump properties: dust clumps with and without associated 6.7-GHz methanol maser emission; dust clumps associated with highly luminous 6.7-GHz methanol masers compared with those of low luminosity; dust clumps associated with 6.7-GHz methanol maser emission with and without associated 12.2-GHz methanol maser emission; and dust clumps with and without associated radio continuum. Our statistics show that there are many differences between the sources that fall into these subgroups; consistent with the different source combinations and characteristics being associated with different stages of massive star formation. An evolutionary scenario for the common maser species associated with high-mass star formation regions is presented.

The positions of the 6.7-GHz methanol masers have been compared to sources in the GLIMPSE point source catalogue. We find that 57 per cent of the 6.7-GHz methanol masers that lie within the GLIMPSE regions are associated with a GLIMPSE point source. Comparison between the GLIMPSE colours associated with sources that exhibit emission at both 6.7- and 12.2-GHz and those exclusively at 6.7-GHz has shown that there is little difference between the colours of the associated point sources. 

We will present detailed investigation of the relative intensities and velocity ranges of 12.2-GHz methanol masers detected towards the 6.7-GHz methanol masers detected in the southern hemisphere component of the MMB survey \citep{Green08} in a subsequent paper. The near simultaneous observations of hundreds of 6.7- and 12.2-GHz methanol masers collected from the MMB survey and follow-up observations will be used to make a detailed investigation of the range and distribution of the
intensity ratio for these two transitions.  Combining this with our
evolutionary studies we will obtain unique insight into the changes in
physical conditions responsible for the presence/absence of the
different maser methanol transitions.

Our results relating to the evolution of star formation regions will be independently tested in the future through comparison between the unbiased sample of 6.7-GHz methanol masers observed in the MMB survey \citep{Green08}, followup observations at 12.2-GHz and 870~$\mu$m dust sources observed as part of ATLASGAL (APEX Telescope Large Area Survey of the Galaxy)\footnote{http://www.mpifr-bonn.mpg.de/div/atlasgal/}.

\section*{Acknowledgments}

We are grateful to Stacy Mader for his help with the observations and with processing data which had been recorded with incorrect time stamps. The Parkes telescope is part of the Australia Telescope which is funded by the Commonwealth of Australia for operation as a National Facility managed by CSIRO. Financial support for this work was provided by the Australian
Research Council. This research has made use of: NASA's Astrophysics
Data System Abstract Service; the NASA/
IPAC Infrared Science Archive (which is operated by the Jet Propulsion
Laboratory, California Institute of Technology, under contract with
the National Aeronautics and Space Administration); the SIMBAD data base, operated at CDS, Strasbourg,
France; and data products from the GLIMPSE
survey, which is a legacy science program of the {\em Spitzer Space
  Telescope}, funded by the National Aeronautics and Space
Administration.

\appendix
\section{Details relating to statistical analysis}\label{appendix}

In this section we present the tables and box plots relating to the statistical analysis that is presented in Section~\ref{Section:stats}. All of the statistical analysis (including the production of box plots) was completed using the R statistical analysis package\footnote{R Development Core Team (2006). R: A language and environment for statistical computing. R Foundation for
  Statistical Computing, Vienna, Austria. ISBN 3-900051-07-0, URL http://www.R-project.org.}. See \citet[][and references therein]{Breen07} for a more comprehensive description of the statistical methods adopted in our analysis.
  
Box plots have been used as a way of presenting the data as they are a useful tool for displaying differences between populations.  They can instantly hilight the distribution of the data as the separation between the different parts of the box have specific meanings. The solid line in the in each of these plots represents the median of the data. The box represents the 25$^{th}$ to the 75$^{th}$ percentile, while the vertical line from the top of the box represents data from the 75$^{th}$ percentile to the maximum value and the vertical line from the bottom of the box represents data from the 25$^{th}$ percentile to the minimum value. Extreme values, or outliers, are represented by dots and are defined as points that exceed 1.5 times the interquartile range in separation from either the 25$^{th}$ or 75${th}$ percentile.

\begin{table}
  \caption{Analysis of deviance table for the single term models; 1.2-mm dust clumps with and without 6.7-GHz methanol}
  \begin{tabular}{ccccc}\hline\label{Tab:6_no}
{\bf Predictor} &         {\bf Deviance}   &  {\bf AIC} &   {\bf LRT}   & {\bf Pr(Chi)}\\  \hline
{\bf none} 			&   455.47 	&457.47   \\                  
{\bf Integrated}    	&   429.24 	&433.24  	&26.23 	&3.034e-07 \\
{\bf Peak}     		&   417.15 	&421.15  	&38.32 	&6.008e-10 \\
{\bf FWHM}     		&   417.91 	&421.91  	&37.56 	&8.881e-10 \\
{\bf Distance}     	&   453.23 	&457.23   	&2.24 	&0.1343038  \\  
{\bf Mass}     		&   445.21 	&449.21  	&10.26 	&0.0013616 \\
{\bf Radius}   		&   443.27 	&447.27  	&12.20 	&0.0004782 \\
{\bf Density}  		&   455.36 	&459.36   	&0.11 	&0.7370057 \\   \hline
	\end{tabular}
\end{table}

\begin{table}
	\caption{Summary table for the Binomial regression model of 1.2-mm dust clumps with and without 6.7-GHz methanol masers, showing for each predictor the estimated coefficient and the standardised z-value and p-value for the test of the hypothesis that $\beta_{i}$=0.}
\begin{tabular}{ccccc}\hline
{\bf Predictor}& {\bf Estimate} & {\bf Std. Error} & {\bf z value} & {\bf p-value}\\ \hline
{\bf Intercept}	& --2.677e+00  	&3.244e-01  	&-8.251  	&$<$ 2e-16\\
{\bf Integrated} 	&--2.952e-01  		&7.338e-02  	&--4.023 	&5.75e-05 \\
{\bf Peak}         	&1.172e+00  		&2.744e-01   	&4.271 	&1.94e-05 \\
{\bf FWHM}       	& 2.994e-02  		&6.427e-03   	&4.659 	&3.18e-06 \\
{\bf Density}      	&1.434e-06  		&6.876e-07   	&2.085   	&0.0370 \\ \hline
         \label{table:6_no_regression}
	\end{tabular}
\end{table}

\begin{figure}
	\psfig{figure=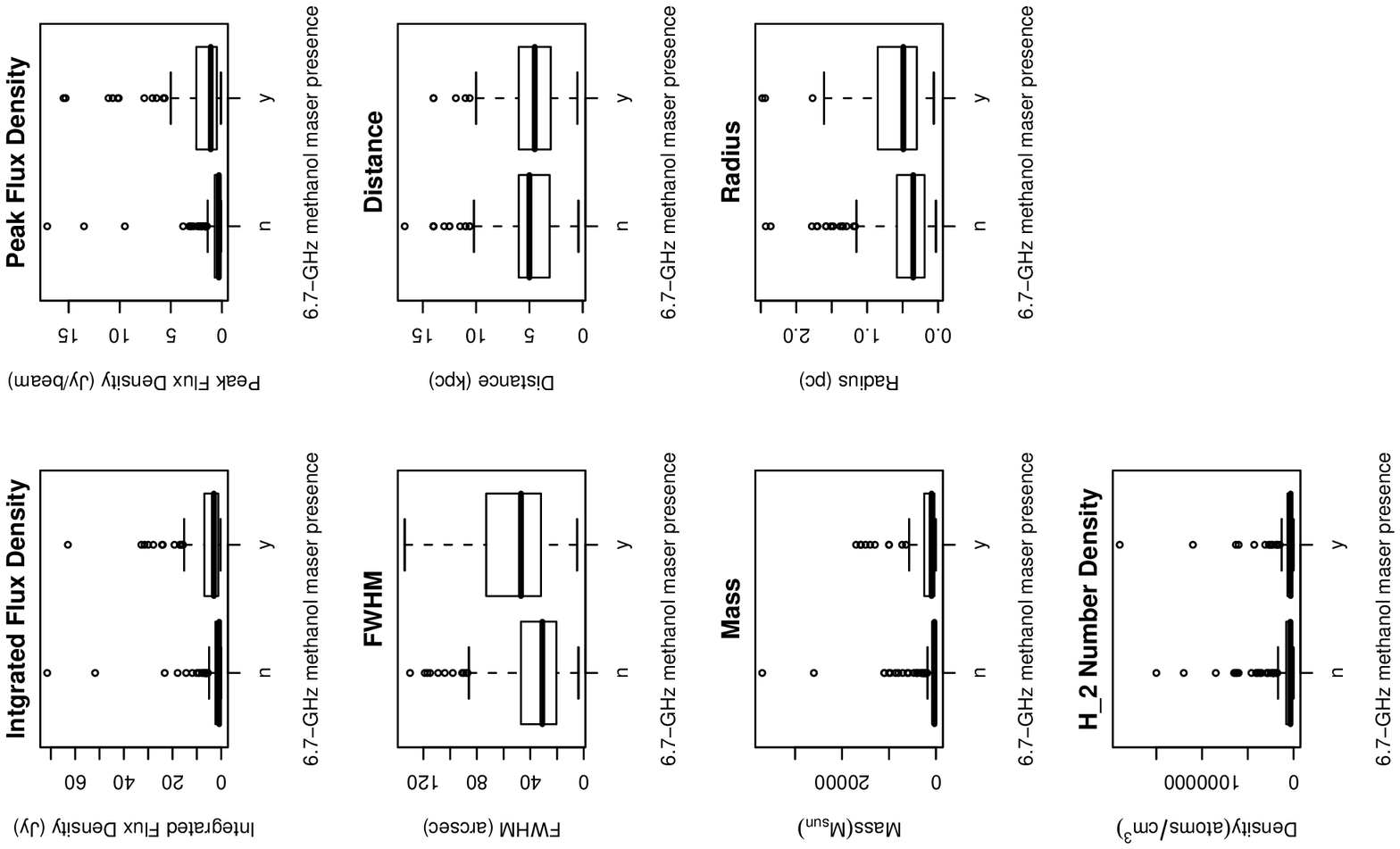,width=14cm,angle=270}
	\caption{Box plots of the 1.2-mm dust clump properties split in the categories of yes and no, referring to the presence or absence of 6.7-GHz methanol maser emission respectively. There is a statistically significant difference between the two categories in the 1.2-mm dust clump integrated flux density, peak flux density, mass, radius and source FWHM. This information is the graphical display relating to the information in Table~\ref{Tab:6_no}.}
  \label{fig:6_no6_box}
\end{figure}

\begin{figure}
	\psfig{figure=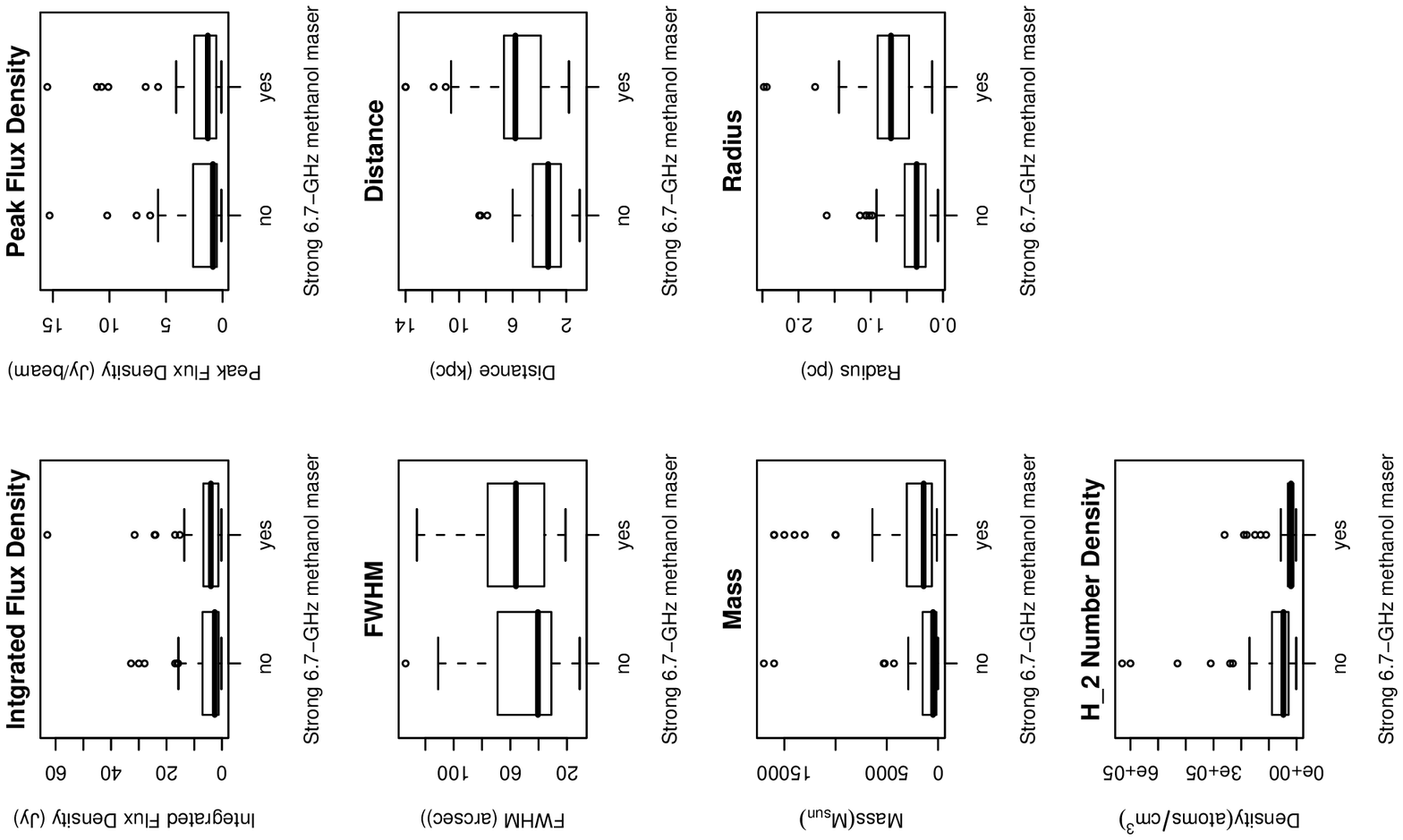,width=13cm,angle=270}
	\caption{Box plots of the 1.2-mm dust clump properties split in the categories of yes and no, referring to the presence of 6.7-GHz methanol maser emission with high and low luminosities respectively. There is a statistically significant difference between the two categories in the 1.2-mm dust clump mass, radius and H$_2$ number density. This information is the graphical display relating to the information in Table~\ref{Tab:strong6weak6}. }
	\label{fig:strong6weak6}
\end{figure}

\begin{table}
  \caption{Analysis of deviance table for the single term models, dust clumps with and without highly luminous 6.7-GHz methanol masers.}
  \begin{tabular}{ccccc}\hline\label{Tab:strong6weak6}
{\bf property} &  {\bf Deviance} &     {\bf AIC}  &   {\bf LRT} & {\bf Pr(Chi)}\\ \hline
{\bf none}      	&140.006 	&142.006          \\            
{\bf Integrated}      	&139.777 	&143.777 		&0.229  	&0.632381   \\  
{\bf Peak}       	&139.690 	&143.690    	&0.316  	&0.574197    \\ 
{\bf FWHM}      	&137.791 	&141.791   	&2.215 	&0.136674  \\    
{\bf Distance}       	&117.717 	&121.717	&22.289 	&2.345e-06\\
{\bf Mass}       	&135.027 	&139.027  	&4.979  	&0.025657\\
{\bf Radius}     	&121.923 	&125.923 	&18.083 	&2.115e-05 \\
{\bf Density}    	&131.441 	&135.441 	&8.564  	&0.003428\\
%{\bf oh}         	&138.873 	&142.873   	&1.054  	&0.304571 \\   
{\bf 6.7 Maser flux}        	&95.731 		&99.731 	&44.275 	&2.854e-11\\ \hline
	\end{tabular}
\end{table}

\begin{table}
	\caption{Summary table for the Binomial regression model of 1.2-mm dust clumps with luminous 6.7-GHz methanol masers and those with less luminous 6.7-GHz methanol masers, showing for each predictor the estimated coefficient and the standardised z-value and p-value for the test of the hypothesis that $\beta_{i}$=0.}
\begin{tabular}{ccccc}\hline
{\bf Predictor}& {\bf Estimate} & {\bf Std. Error} & {\bf z value} & {\bf p-value}\\ \hline
{\bf Intercept} 	&-3.183923  &0.823230  	&-3.868 	&0.00011 	\\
{\bf FWHM}    	&0.017130   	&0.008611   	&1.989 	&	0.04665 \\
{\bf Distance}  	&0.488554   	&0.121787   	&4.012 	&6.03e-05\\ \hline
         \label{table:s6_w6_regression}
	\end{tabular}
\end{table}

\begin{table}
  \caption{Analysis of deviance table for the single term models, dust clumps with 6.7-GHz methanol masers with and without 12.2-GHz methanol. This analysis includes not only the 1.2-mm dust clump properties, but also both the peak flux density and luminosity of the associated 6.7-GHz methanol maser.} 
  \begin{tabular}{ccccc}\hline\label{Tab:612vs6only}
{\bf Predictor} &         {\bf Deviance}   &  {\bf AIC} &   {\bf LRT}   & {\bf Pr(Chi)}\\   \hline    
{\bf none}     		&133.763 &	&135.763     \\                  
{\bf Integrated}     	&133.345 &137.345   &0.417   &0.51821\\
{\bf Peak}   		&133.342 &137.342   &0.421   &0.51655   \\
{\bf FWHM}     		&131.321 &135.321   &2.442   &0.11814   \\
{\bf Distance}     	&133.588 &137.588   &0.175   &0.67568 \\
{\bf Mass}     		&133.559 & 137.558   &0.205   &0.65090  \\  
{\bf Radius}   		&133.573 &137.573   &1.190   &0.27529\\
{\bf Density}  		&130.732 &134.732   &3.031   &0.08171 \\
{\bf 6.7-GHz lum}    	& 113.790 &117.790  &19.973 &7.853e-06\\
{\bf 6.7-GHz flux}   	& 107.800 &111.800  &25.963 &3.481e-07  \\ \hline
	\end{tabular}
\end{table}

\begin{figure}
	\psfig{figure=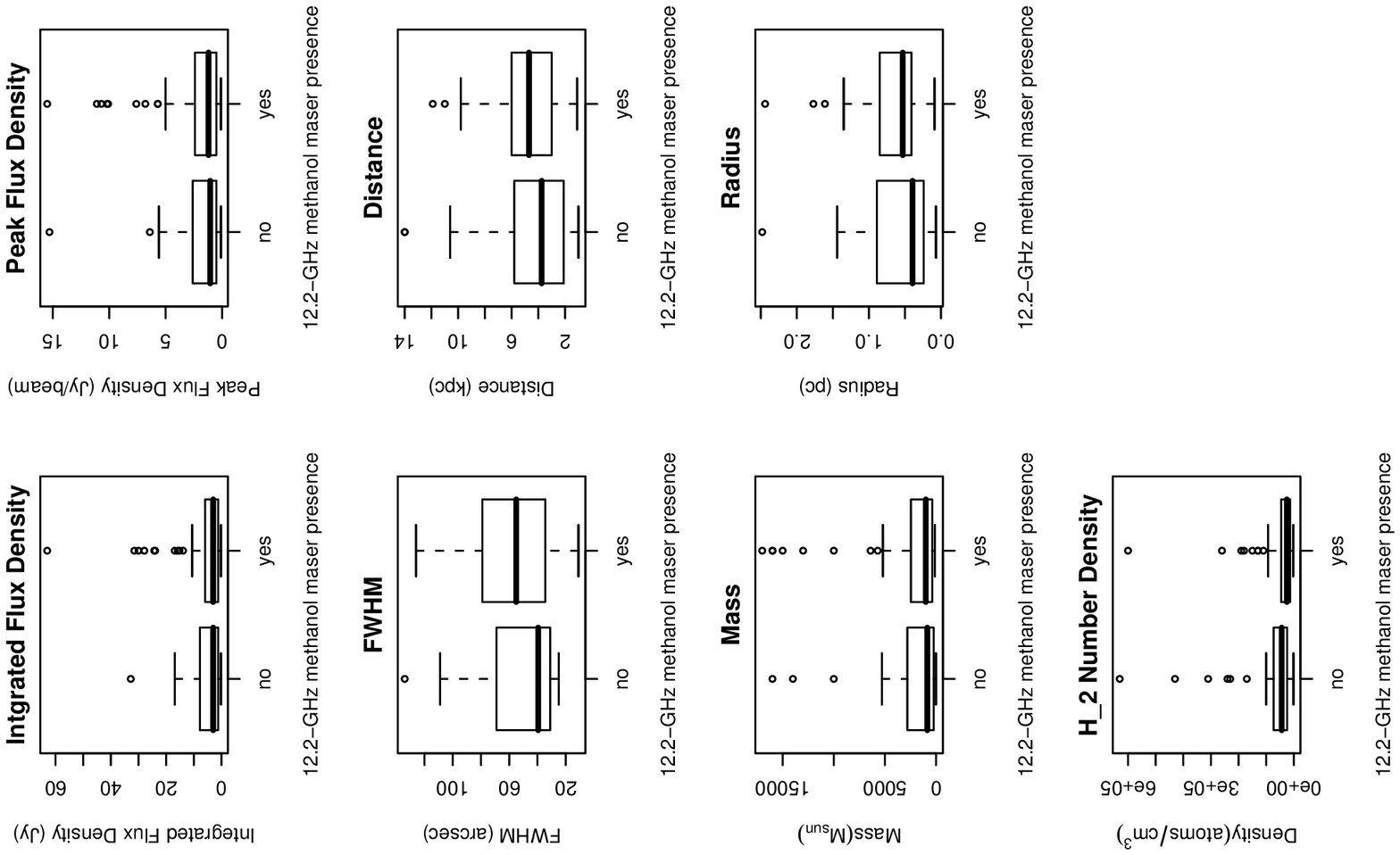,width=13cm,angle=270}
	\caption{Box plots of the 1.2-mm dust clump properties split in the categories of yes and no, referring to the presence or absence of 12.2-GHz methanol maser emission. There is a statistically significant difference between the two categories in the 1.2-mm dust clump H$_2$ number density. This information is the graphical display relating to some of the information in Table~\ref{Tab:612vs6only}.}
	\label{fig:6_6_12}
\end{figure}

\begin{figure}
	\psfig{figure=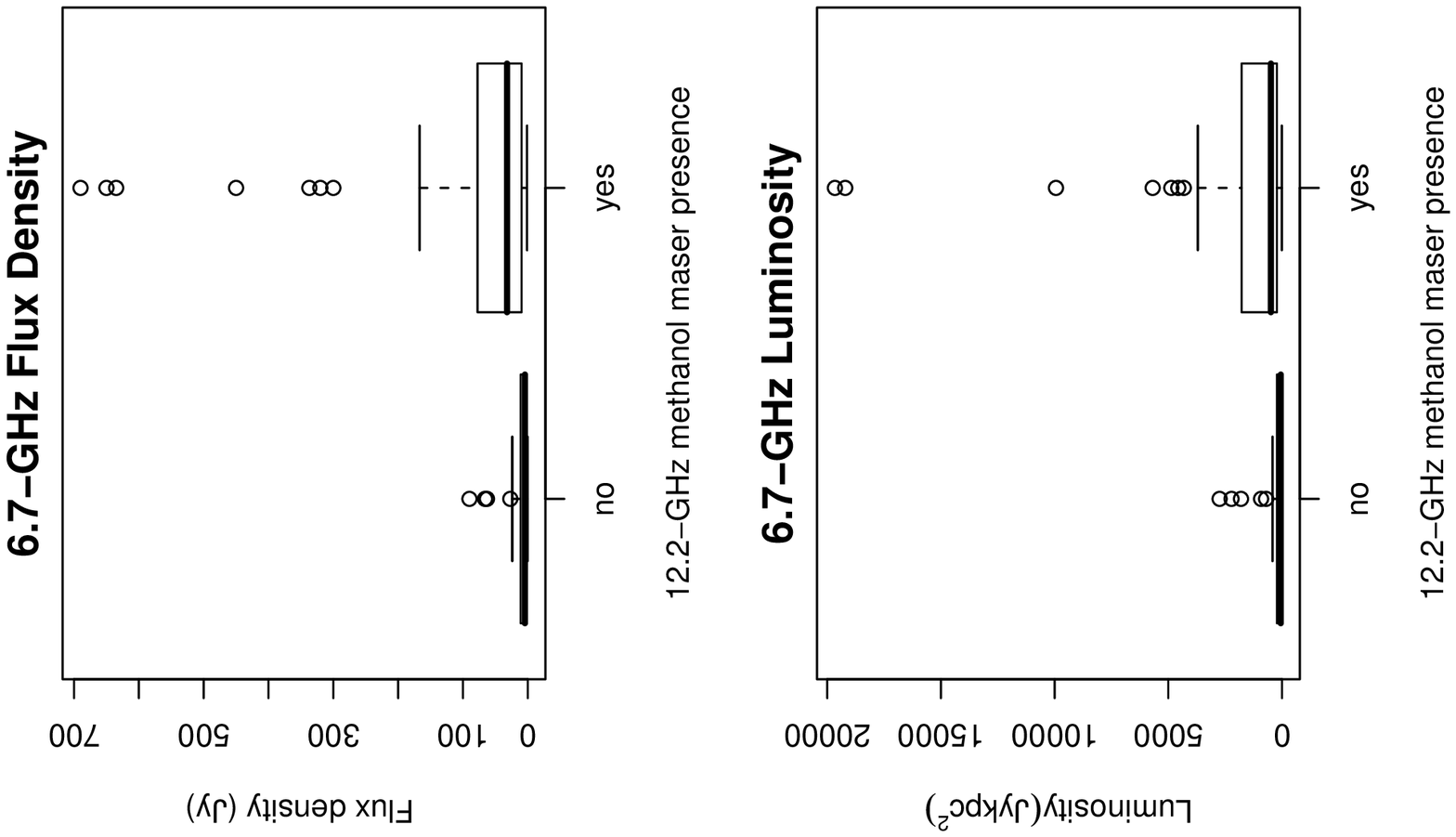,width=10cm,height=8cm,angle=270}
	\caption{Box plots of the 6.7-GHz methanol maser properties split in the categories of yes and no, referring to the presence or absence of 12.2-GHz methanol maser emission. The differences in the two populations is highly significant in the case of both the 6.7-GHz methanol maser peak flux density and peak luminosity. The three points with the most extreme values for either property have been removed to allow the distribution to be seen. This information is the graphical display relating to some of the information in Table~\ref{Tab:612vs6only}.}
\label{fig:6_6_12flux}
\end{figure}

\begin{table}
	\caption{Summary table for the Binomial regression model of 1.2-mm dust clumps with associated 6.7-GHz methanol masers with and without a 12.2-GHz methanol maser counterpart, showing for each predictor the estimated coefficient and the standardised z-value and p-value for the test of the hypothesis that $\beta_{i}$=0.}
\begin{tabular}{ccccc}\hline
{\bf Predictor}& {\bf Estimate} & {\bf Std. Error} & {\bf z value} & {\bf p-value}\\ \hline
{\bf Intercept} 	&-0.41058    &0.29388  &-1.397  &0.16238   	\\
{\bf 6.7-GHz flux}  	&0.03571  &  0.01180  & 3.027 & 0.00247\\ \hline
         \label{table:6_6_12_regression}
	\end{tabular}
\end{table}

\begin{table}
  \caption{Analysis of deviance table for the single term models, dust clumps with 6.7-GHz methanol masers with and without 12.2-GHz methanol. Data associated with 6.7-GHz with peak flux densities less than 9~Jy have been removed. The results of the single term models applied to the full data set are shown in Table~\ref{Tab:612vs6only}. }
  \begin{tabular}{ccccc}\hline\label{Tab:612vs6allmorethan45Jy}
{\bf Predictor} &         {\bf Deviance}   &  {\bf AIC} &   {\bf LRT}   & {\bf Pr(Chi)}\\   \hline    
{\bf none}     		& 60.925 	&62.925 \\
{\bf Integrated}     	&60.428 	&64.428  	&0.496 	&0.481075   \\    
{\bf Peak}   		& 59.637 	&63.637  	&1.287 	&0.256532    \\
{\bf FWHM}     		& 59.856 	&63.856  	&1.069 	&0.301177   \\
{\bf Distance}     	& 59.343 	&63.343  	&1.582 	&0.208483   \\
{\bf Mass}     		& 60.924 	&64.924  	&0.0003 	&0.986207    \\
{\bf Radius}   		& 60.342 	&64.342  	&0.581 	&0.445201   \\
{\bf Density}  		& 55.748 	&59.748  	&5.177 	&0.022894 \\
{\bf 6.7-GHz Lum}    	& 55.136	&59.136  	&5.789 	&0.016130 \\
{\bf 6.7-GHz Flux} 	& 53.484 	&57.484  	&7.440 	&0.006377\\ \hline

%oh       1   59.139 63.139  1.351 0.245092   
	\end{tabular}
\end{table}

\begin{table}
  \caption{Analysis of deviance table for the single term models, dust clumps with and without associated 8-GHz radio continuum.}
  \begin{tabular}{ccccc}\hline\label{Tab:radiocont}
{\bf Predictor} &         {\bf Deviance}   &  {\bf AIC} &   {\bf LRT}   & {\bf Pr(Chi)}\\   \hline    
{\bf none}     		&  365.19 &367.19  \\
{\bf Integrated}     	& 340.56 &344.56  &24.63 &6.941e-07\\    
{\bf Peak}   		&  330.03 &334.03  &35.17 &3.025e-09   \\
{\bf FWHM}     		& 305.12 & 309.12  &60.07 &9.145e-15   \\
{\bf Distance}     	& 363.24 & 367.24   &1.96  &0.161618    \\
{\bf Mass}     		& 339.94 & 343.94  &25.26 &5.023e-07  \\
{\bf Radius}   		& 320.32 &324.32  &44.88 &2.096e-11  \\
{\bf Density}  		&355.57 &359.57   &9.63  &0.001919  \\ \hline
	\end{tabular}
\end{table}

\begin{figure}
	\psfig{figure=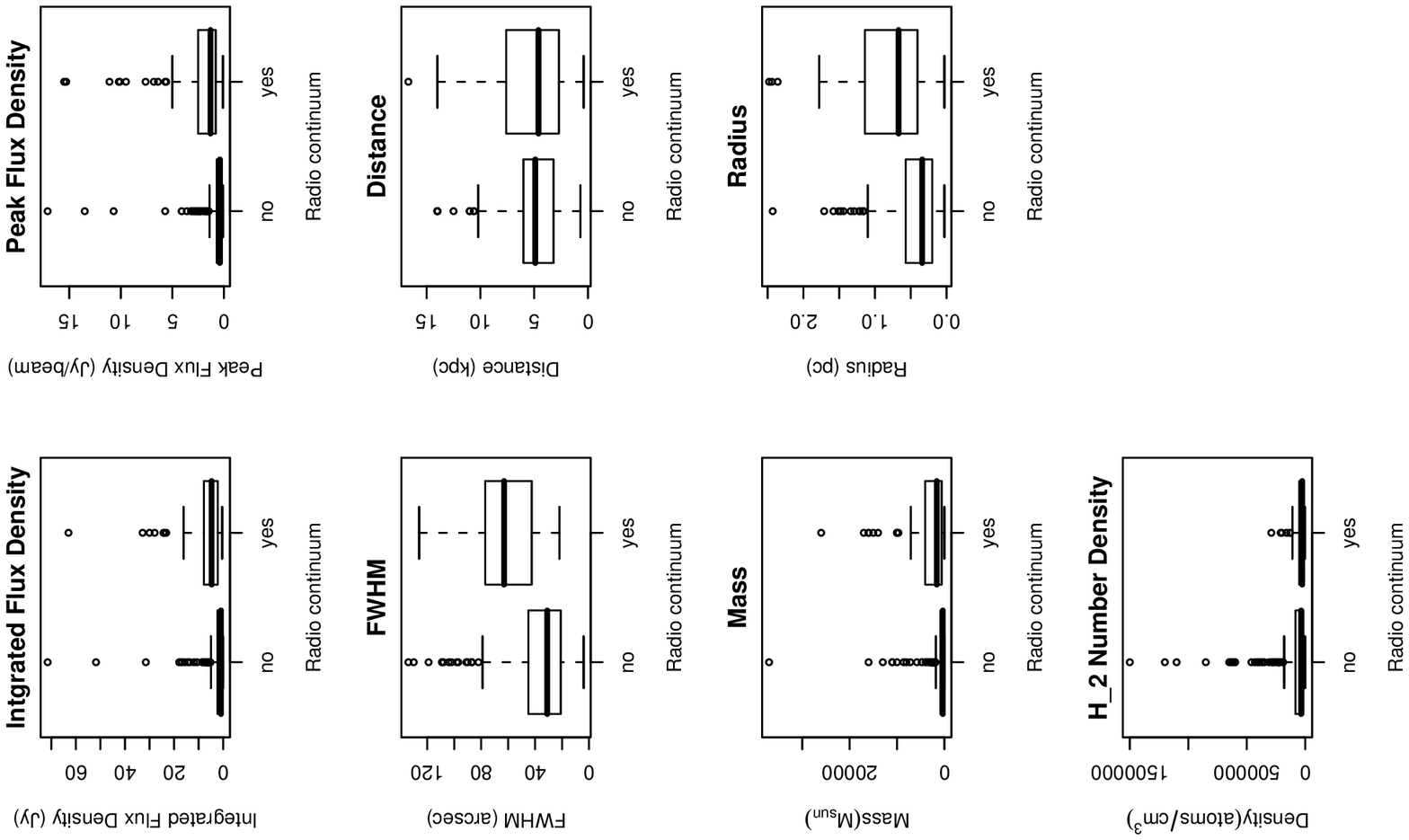,width=13cm,angle=270}
	\caption{Box plots of the 1.2-mm dust clump properties split in the categories of yes and no, referring to the presence or absence of detectable radio continuum emission. There is a statistically significant difference between the two categories in all 1.2-mm dust clump properties with the exception of distance.  The results of the single term models applied to the full data set are shown in Table~\ref{Tab:radiocont}.}
	\label{fig:radio}
\end{figure}

\end{document}